\documentclass{appolb}
\usepackage{graphicx}
\usepackage{amsmath,amssymb}
\usepackage{cite}
 
\newcommand{\MSb}{\overline{\mathrm{MS}}}
\let\oldbibliography\thebibliography
\renewcommand{\thebibliography}[1]{\oldbibliography{#1}
\setlength{\baselineskip}{0pt}
\setlength{\itemsep}{2pt}} 

\begin{document}
\title{Generalized Parton Distributions from Lattice QCD%
\thanks{Presented by K.\ Cichy at the Cracow Epiphany Conference 2023.}%
}
\author{First Author, second Author
\address{affiliation}
\\[3mm]
{Third Author 
\address{affiliation}
}
\\[3mm]
the Name(s) of other Author(s)
\address{affiliation}
}

\author{Krzysztof Cichy$^a$, Shohini Bhattacharya$^b$, Martha Constantinou$^c$, Jack Dodson$^c$, Xiang Gao$^d$, Andreas Metz$^c$, Joshua Miller$^c$, Swagato Mukherjee$^e$, Aurora Scapellato$^c$, Fernanda Steffens$^f$, Yong Zhao$^d$
\address{\footnotesize $^a$ Faculty of Physics, Adam Mickiewicz University, \\ ul.\ Uniwersytetu Pozna\'nskiego 2, 61-614 Pozna\'{n}, Poland
\\$^b$ RIKEN BNL Research Center, Brookhaven National Laboratory, \\ Upton, NY 11973, USA
\\$^c$ Department of Physics,  Temple University,  Philadelphia,  PA 19122 - 1801,  USA
\\$^d$ Physics Division, Argonne National Laboratory, Lemont, IL 60439, USA
\\$^e$ Physics Department, Brookhaven National Laboratory, Upton, NY 11973, USA
\\$^f$ Institut f\"ur Strahlen- und Kernphysik, Rheinische Friedrich-Wilhelms-Universit\"at, Nussallee 14-16, 53115 Bonn}
}

\maketitle
\begin{abstract}
In recent years, there has been a breakthrough in lattice calculations of $x$-dependent partonic distributions. This encompasses also distributions describing the 3D structure of the nucleon, such as generalized parton distributions (GPDs). We report a new method of accessing GPDs in asymmetric frames of reference, relying on a novel Lorentz-covariant parametrization of the accessed off-forward matrix elements in boosted nucleon states. The approach offers the possibility of computationally more efficient determination of the full parameter dependence of GPDs and as such, it can contribute to better understanding of nucleon's structure.
\end{abstract}
  
\section{Introduction}
One of the main aims of hadronic physics for the next years is to better understand the internal structure of the nucleon.
This concerns, in particular, its three-dimensional structure -- namely, how the quarks and gluons inside of the nucleon move and how they are distributed in the plane transverse to the nucleon's direction of motion. Understanding these details will help us, for example, to answer fundamental questions about the origin of the nucleon's mass and spin. The prospects for answering these questions are very robust in the light of the currently ongoing experiments, such as the 12 GeV program at the Jefferson Lab  \cite{Burkert:2018nvj} and, particularly, the Electron-Ion Collider under construction at BNL \cite{AbdulKhalek:2021gbh}.
However, true progress will require also various theoretical contributions.
One of them will be the determination of different partonic distributions from first principles, using lattice quantum chromodynamics (lattice QCD).

The three-dimensional structure is described by generalized parton distributions (GPDs) and transverse-momentum-dependent parton distribution functions (TMDs), the generalizations of PDFs to non-forward kinematics and unintegrated transverse momenta, respectively.
Some of these functions boil down to PDFs in a certain limit, but they contain much more information about the nucleon's structure.
In this proceeding, we concentrate on GPDs, for which significant progress has been achieved recently.

There is no direct access to any partonic distributions on the lattice, since the underlying theory, QCD, is discretized in Euclidean spacetime.
However, the situation is largely analogous to accessing such distributions from experiment -- there are no direct experimental measurements either.
Instead, observables can be factorized into partonic functions and perturbative parts.
In the same spirit, one can devise lattice observables factorizable into PDFs/GPDs/TMDs via a perturbative ``matching'' kernel.
Early attempts to do so date back to around 20 years ago, but a real breakthrough was the arrival of the seminal proposal of Ji in 2013 \cite{Ji:2013dva,Ji:2014gla}.
There, analogues of partonic functions were formulated, with Minkowski-space light-front correlations replaced by lattice-calculable (Euclidean-space) spatial ones for a boosted hadron.
This defines the so-called quasi-distributions, which approach their light-cone counterparts for infinite hadron momentum.
However, clearly, actual simulations can only utilize finite hadron momenta. But the finite-boost quasi-distributions can be perturbatively related to partonic ones due to their difference being only in the ultraviolet regime.

The progress in lattice calculations of partonic distributions has been impressive in the last decade and for its account, we refer to several existing reviews~\cite{Cichy:2018mum,Ji:2020ect,Constantinou:2020pek,Cichy:2021lih,Cichy:2021ewm}.
Most of the work described there concern PDFs, which have been the natural starting points for lattice evaluations with modern methods.
The applications to GPDs, in turn, have been comparatively limited, see e.g.~Refs.\ \cite{Ji:2015qla, Xiong:2015nua, Bhattacharya:2018zxi, Liu:2019urm, Bhattacharya:2019cme, Chen:2019lcm, Radyushkin:2019owq, Ma:2019agv, Luo:2020yqj, Alexandrou:2020zbe,Alexandrou:2021bbo,CSSMQCDSFUKQCD:2021lkf,Bhattacharya:2021oyr, Ma:2022ggj, Shastry:2022obb, Ma:2022gty,Ji:2022thb}.
The relatively less advanced status of lattice GPD calculations naturally reflects that they are more difficult objects to determine than PDFs.
To a large extent, this is due to the fact that GPDs depend on more variables than just the parton momentum fraction $x$, i.e.\  the total momentum transfer $t$ and the longitudinal momentum transfer (skewness, $\xi$).
Until recently, a major hindrance in lattice calculations of GPDs has been the necessity of using the symmetric frame of reference, in which the momentum transfer is symmetrically distributed between the initial and final state.
Consequently, the all-to-all propagator required for the determination of the appropriate matrix elements had to be separately computed for each value of $t$.
Last year, we considered for the first time asymmetric frames of reference for the computation of GPDs.
In such a setup, where the whole momentum transfer is ascribed to the source state, one can access several values of $t$ from a single calculation with a fixed sink state of zero transverse momentum.
Below, we summarize the method and our procedure and we show our proof-of-concept calculation that demonstrates the feasibility and correctness of the approach.

\vspace*{-2mm}
\section{GPDs in different frames of reference}
Quasi-GPDs are defined from the matrix elements (MEs) of a nonlocal operator,
$\langle N(P_f)|\bar\psi\left(z\right)\Gamma W(0,z)\psi\left(0\right)|N(P_i)\rangle$,
where $\Gamma$ is the Dirac structure appropriate for the targeted type of GPDs, $|N(P_{i/f})\rangle$ is the initial/final state boosted to momentum $P_{i/f}$ and $W(0,z)$ is a Wilson line of length $z$, taken in the $z$-direction. We also define the momentum transfer 4-vector, $\Delta=P_f-P_i$.
These MEs, that we denote by $F_{\Gamma}(z,P,\Delta)$, can be obtained from a suitable ratio of two-point and three-point correlation functions.
According to Ref.~\cite{Bhattacharya:2022aob}, to which we refer for more details, the MEs for the vector case ($\Gamma=\gamma^\mu$, $\mu=0,1,2,3$) can be parametrized in terms of eight linearly-independent Lorentz structures, 
\begin{align}
\label{eq:parametrization_general}
F^{\mu}_\Gamma (z,P,\Delta) & = \bar{u}(p_f,\lambda') \bigg [ \dfrac{P^{\mu}}{m} A_1 + m z^{\mu} A_2 + \dfrac{\Delta^{\mu}}{m} A_3 + i m \sigma^{\mu z} A_4 + \dfrac{i\sigma^{\mu \Delta}}{m} A_5 \nonumber \\
& \hspace{1cm} + \dfrac{P^{\mu} i\sigma^{z \Delta}}{m} A_6 + m z^{\mu} i\sigma^{z \Delta} A_7 + \dfrac{\Delta^{\mu} i\sigma^{z \Delta}}{m} A_8  \bigg ] u(p_i, \lambda),
\end{align}
where $A_i$ are Lorentz-invariant amplitudes.

We consider here the zero-skewness case ($\xi=0$), i.e.\ $\vec{\Delta}=(\Delta_1,\Delta_2,0)$. The standard symmetric (Breit) frame has $\vec{P}^{\,s}_i = \vec{P} - {\vec{\Delta}}/2,\,\vec{P}^{s}_f = \vec{P} + {\vec{\Delta}}/2$, and the asymmetric one $\vec{P}^{\,a}_i = \vec{P} - {\vec{\Delta}}, \, \vec{P}^{a}_f = \vec{P}$.
All frame-dependent expression will be written with an upper index $s/a$ (symmetric/asymmetric).

We denote the parity-projected Euclidean MEs by $\Pi_\mu(\Gamma_\kappa)$ (where $\Gamma_0=(1+\gamma_0)/4$ (unpolarized projector), $\Gamma_k=(1+\gamma_0) i \gamma_5 \gamma_k/4$ (polarized projector, $k=1,2,3$)) and below, we give example expressions for the $\gamma_0$ insertion and the unpolarized projector:
\begin{align}
\label{eq:P0G0sym}
\Pi^s_0(\Gamma_0) & = C \Bigg( \frac{ E  \left( E   ( E  +m)- P_3^2\right)}{2 m^3} A_1
    +\frac{    ( E  +m)    \left(-E ^2+m^2+ P_3^2\right)}{m^3} A_5 \nonumber\\
    &\hspace{1.8cm} +\frac{  E   P_3  \left(-E ^2+m^2+ P_3^2\right)   z}{m^3} A_6
  \Bigg),
\end{align}
\begin{align}
\label{eq:P0G0asym}
\Pi^a_0(\Gamma_0) &= C  \Bigg(
 -\frac{  E^+ ( E^- -2 m) ( E_f +m)}{8 m^3}A_1
    -\frac{ E^- ( E^- -2 m) ( E_f +m) }{4m^3} A_3 \nonumber\\
     &\hspace{11mm}-\frac{  E^-  P_3  z}{4 m} A_4
    +\frac{ E^+ E^- ( E_f +m) }{4 m^3} A_5
     +\frac{  E^+ E^- E_f P_3 z}{4 m^3}A_6\nonumber\\
    &\hspace{11mm}+ \frac{ (E^-)^2 E_f P_3 z }{2 m^3} A_8
  \Bigg), 
\end{align}
where $C= 2 m^2 / \sqrt{E_f E_i (E_f + m) (E_i + m)}$, $E_{i/f}=\sqrt{\vec{P}_{i/f}^2+m^2}$, $E\equiv E_f=E_i$ (symmetric frame), $E^\pm=E_f\pm E_i$ (asymmetric frame) and $m$ is the nucleon mass.
These explicit expressions show that the MEs are frame-dependent and, in general, more complicated in the asymmetric frame.
In the above example, the MEs in both frames receive contributions from the amplitudes $A_1,A_5,A_6$, but the kinematics of the asymmetric frame also induces contributions from $A_3,A_4,A_8$ and modifies the kinematic coefficients of $A_1,A_5,A_6$.
We emphasize that while MEs depend on the chosen kinematics, the amplitudes are Lorentz-invariant and below, we demonstrate this for our results.

Once the amplitudes are extracted from either frame, they can be used to calculate the $H$ and $E$ (coordinate-space) GPDs. The standard definition of these functions, employing only the $\gamma_0$ insertion in the three-point correlator, leads to frame-dependent expressions for the GPDs:
\begin{equation}
\label{eq:Hsym}
H^s(A_i)   =  
A_1 + \frac{\Delta_\perp^2 z}{2 P_3}  {{A_6}},
\end{equation}
\begin{equation}
\label{eq:Esym}
E^s(A_i)   =  
- A_1  + 2 A_5 + \frac{\left( 4 E^2 - \Delta_\perp^2\right) z}{2P_3}   {{A_6}},
\end{equation}
\begin{align}
\label{eq:Hasym}
H^a(A_i) & = 
A_1 + \frac{2E^-}{E^+} {{A_3}} + \frac{E^- m^2 z}{E^+ P_3}  {{A_4}}+ \frac{ (\Delta_\perp^2-(E^-)^2)z}{2 P_3}  {{A_6}} \nonumber\\
&+ \frac{E^- (\Delta_\perp^2-(E^-)^2)z}{E^+ P_3}   {{A_8}},
\end{align}
\begin{align}
\label{eq:Easym}
E^a(A_i)  & = 
- A_1 - \frac{2E^-}{E^+}  {{A_3}} - \frac{2m^2 E_f z}{E^+P_3}  {{A_4}} + 2 A_5 \\ 
& + \frac{\left(4E_f^2 - E^+ E^- - \Delta_\perp^2 \right)z}{2P_3}   {{A_6}} 
  + \frac{z E^- \left(4E_f^2 - E^+ E^- - \Delta_\perp^2 \right)z}{E^+ P_3}   {{A_8}},\nonumber
\end{align}
where $\Delta_\perp^2=\Delta_1^2 + \Delta^2_2$.
However, one can also introduce another definition for the quasi-GPDs that is made Lorentz-invariant (see Ref.~\cite{Bhattacharya:2022aob} for more details).
We will denote it with $H/E$ without a frame index: 
\begin{equation}
\label{eq:HLI}
H(A_i) =  A_1 ,
\end{equation}
\begin{equation}
\label{eq:ELI}
E(A_i) = - A_1 + 2 A_5 + 2 z P_3 A_6 .
\end{equation}
In these expressions, there are no contributions from $A_3, A_4, A_8$ and the contribution from $A_6$ is either removed ($H$-function) or reduced ($E$-function).
In terms of operator insertions, they correspond to adding contributions from the $\gamma_1$ and $\gamma_2$ Dirac structures, i.e.\ one formally defines a new operator with additional contributions.
However, we emphasize that this new operator for evaluating quasi-GPDs leads to the same light-cone limit, with modified convergence properties, including a possibly faster convergence to light-cone GPDs.
Below, we show a first test of these convergence properties for both $H$ and $E$.

\section{Procedure to extract $x$-dependent light-cone GPDs}
\label{sec:procedure}
We outline here the full procedure to get from lattice-calculated observables to light-cone GPDs.

\noindent\textbf{1.\ Computation of bare MEs.} 
The first step is to calculate bare MEs of the relevant operators using the chosen frame of reference.
This is the most costly part of the procedure and it involves separate computations for every sink momentum.
As discussed above, this boils down to independent calculations for each $t$ in the symmetric frame, where the sink momentum depends explicitly on $\vec{\Delta}$.
In the asymmetric frame, in turn, one can work with a fixed sink momentum $(0,0,P_3)$ and obtain MEs for different $t$ within a single calculation.
This constitutes the main advantage of the asymmetric frame.
For the standard definition of GPDs, one only needs the MEs of the $\gamma_0$ operator, i.e.\ $\Pi_0(\Gamma_{0,1,2})$.
To utilize the Lorentz-invariant definition, additional MEs are computed with the $\gamma_1$, $\gamma_2$ insertions.\\
\noindent\textbf{2.\ Extraction of the amplitudes and calculation of $H/E$ in coordinate space.}
Given the lattice-evaluated MEs, the amplitudes $A_i$ can be extracted by solving the linear system of equations that relate $\Pi_\mu(\Gamma_\kappa)$ to $A_i$'s.
The quasi-$H/E$ functions can be expressed in terms of these amplitudes (or directly in terms of MEs, i.e.\ the amplitudes do not need to be explicitly known).\\
\noindent\textbf{3.\ Non-perturbative renormalization of $H/E$.}
The outcome of the previous step are bare quasi-$H/E$ functions in coordinate space that contain standard logarithmic and Wilson-line-induced power divergences.
For this work, we choose to renormalize them with a variant of the regularization-independent momentum subtraction scheme (RI/MOM) \cite{Constantinou:2017sej,Alexandrou:2017huk}.
We note that alternative renormalization procedures are being investigated, such as the hybrid scheme \cite{Ji:2020brr}.\\
\noindent\textbf{4.\ Reconstruction of $x$-dependence.}
The renormalized coordinate-space quasi-GPDs are then brought into momentum (Bjorken-$x$) space.
This is a non-trivial and currently ambiguous step, given that one attempts to reconstruct a continuous distribution from a finite set of lattice evaluations, additionally truncated at some finite value of the Wilson line length.
Mathematically, this poses an ill-defined inverse problem \cite{Karpie:2019eiq}.
In this work, we apply the Backus-Gilbert method \cite{BackusGilbert} as implemented in Ref.~\cite{Bhat:2020ktg}.
The inverse problem is tackled by an additional model-independent assumption that selects the one distribution that has the minimal variance with respect to the statistical variation of the input data.
Clearly, this provides a formal solution to the inverse problem in the sense of reducing the number of $x$-space distributions corresponding to the discrete data from infinity to one, but a true solution for the future is to provide the actual data that is missing.
While it is not possible to get continuous data from the lattice, significantly denser data will milden the problem to a sufficient extent.\\
\noindent\textbf{5.\ Matching to light-cone GPDs.}
The final step is to translate the $x$-dependent quasi-GPD from Euclidean to Minkowski spacetime, yielding a corresponding light-cone GPD.
The factorization formula that relates the quasi-GPD to the light-cone GPD via a perturbative matching kernel is valid up to power-suppressed corrections in $1/P_3^2$.
Thus, the momentum $P_3$ has to be as large as possible.
In practice, the access to large values of $P_3\gtrsim2-3$ GeV is impossible, due to the statistical signal exponentially decaying with increasing boost.
We use the matching formulae derived in Ref.~\cite{Liu:2019urm} that connect the RI/MOM-renormalized quasi-GPD to the light-cone GPD in the $\MSb$ scheme and we evaluate the GPDs at the scale 2 GeV.

\section{Results}
All the results in this section were obtained from a single lattice ensemble in a setup characterized by:
\begin{itemize}
\itemsep-1mm
\item $N_f=2+1+1$ twisted-clover quarks with Iwasaki gluons \cite{Alexandrou:2018egz},
\item lattice spacing of $a\approx0.093$ fm,
\item $32^3\times64$ lattice (physical extent $L\approx3$ fm),
\item non-physical pion mass $m_\pi\approx260$ MeV,
\item source-sink separation in the three-point functions of $t_s=10a$,
\item nucleon momentum in the $z$-direction: $P_3=0.83, 1.25, 1.67$ GeV,
\item momentum transfers varying from $-t=0.17$ to $2.24$ GeV$^2$.
\end{itemize}

\begin{figure}[t!]
\centering
\includegraphics[angle=270,width=6.8cm]{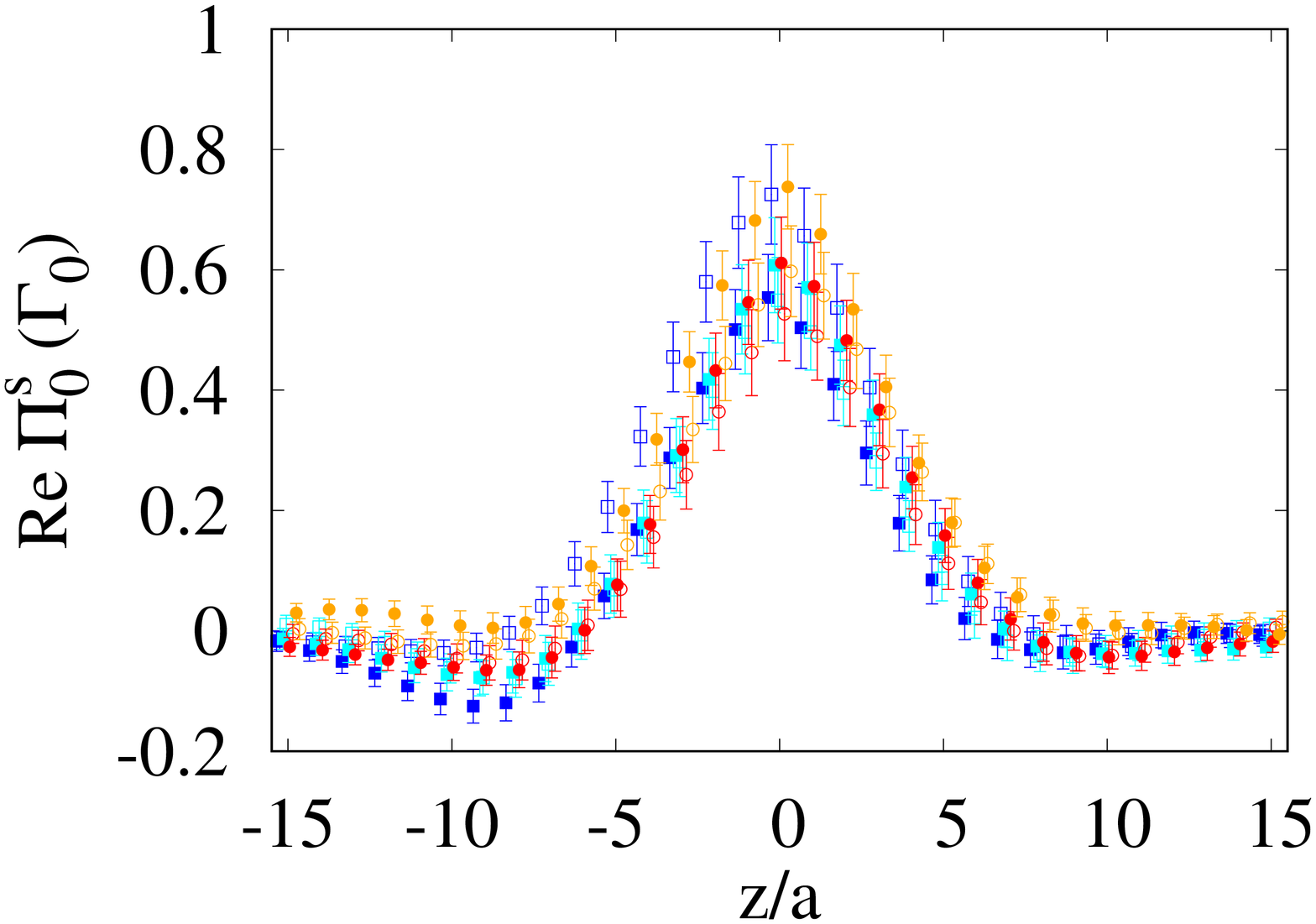}\hspace*{-5mm}
\includegraphics[angle=270,width=6.8cm]{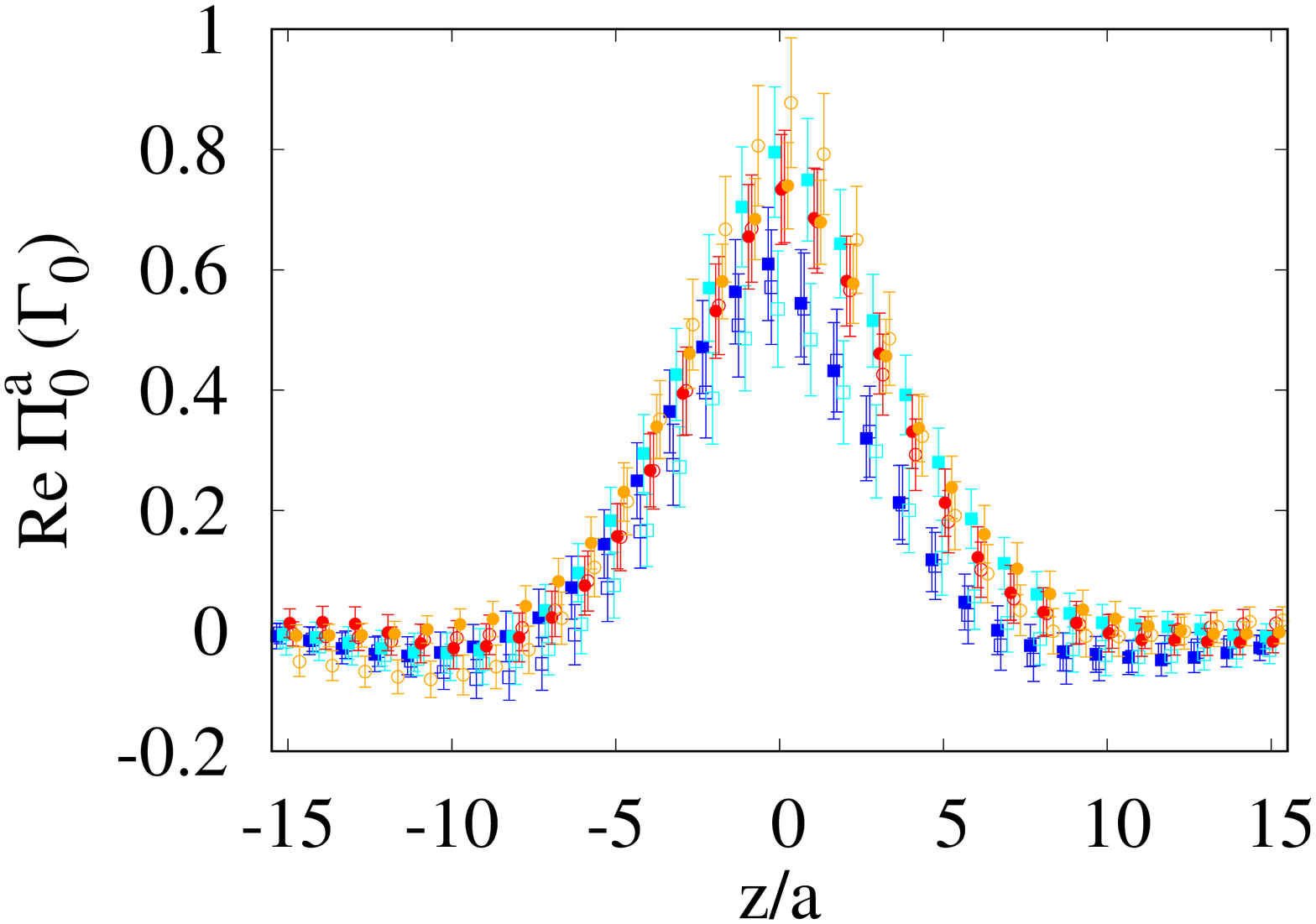}\vspace*{-3mm}
\includegraphics[angle=270,width=6.8cm]{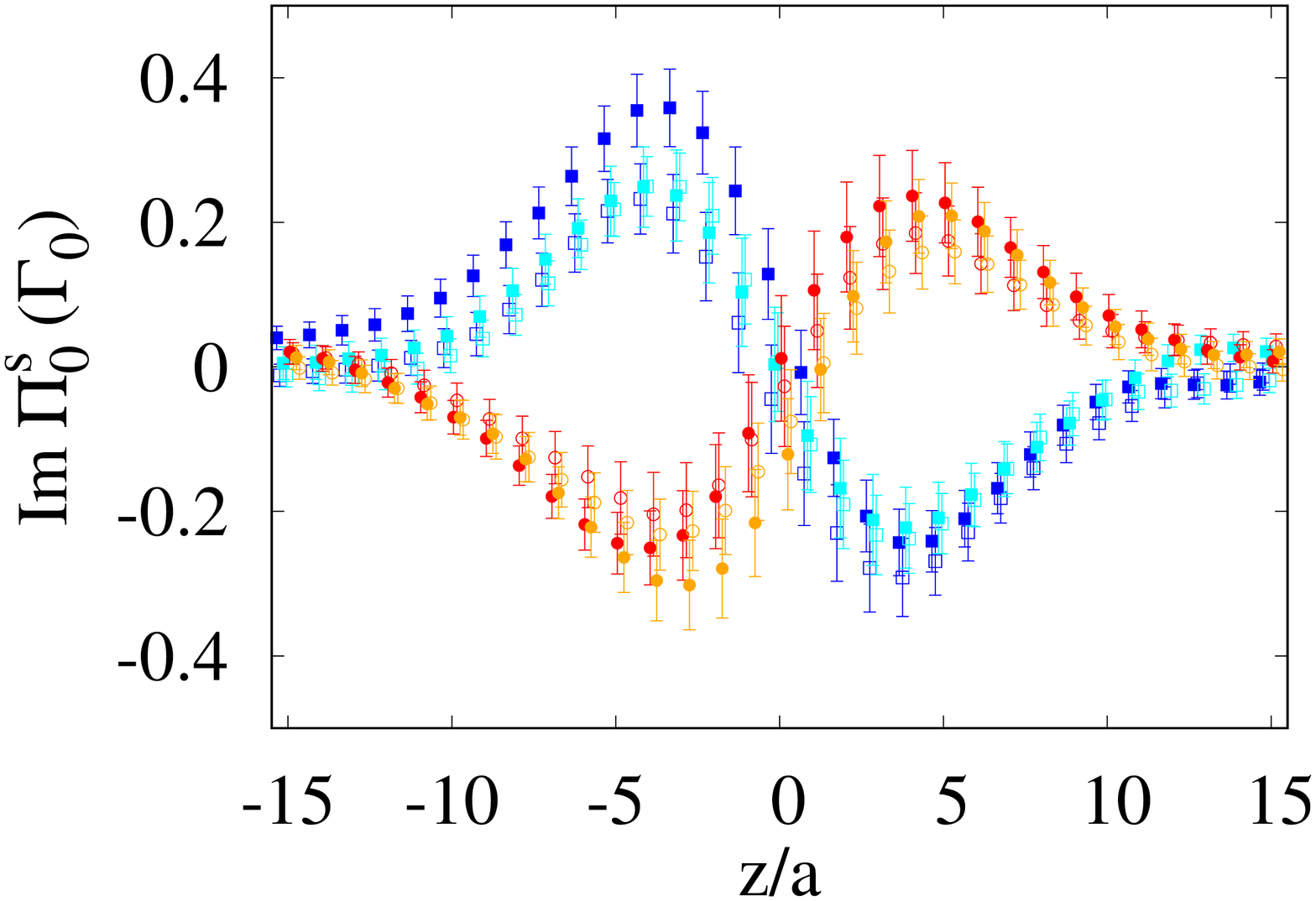}\hspace*{-5mm}
\includegraphics[angle=270,width=6.8cm]{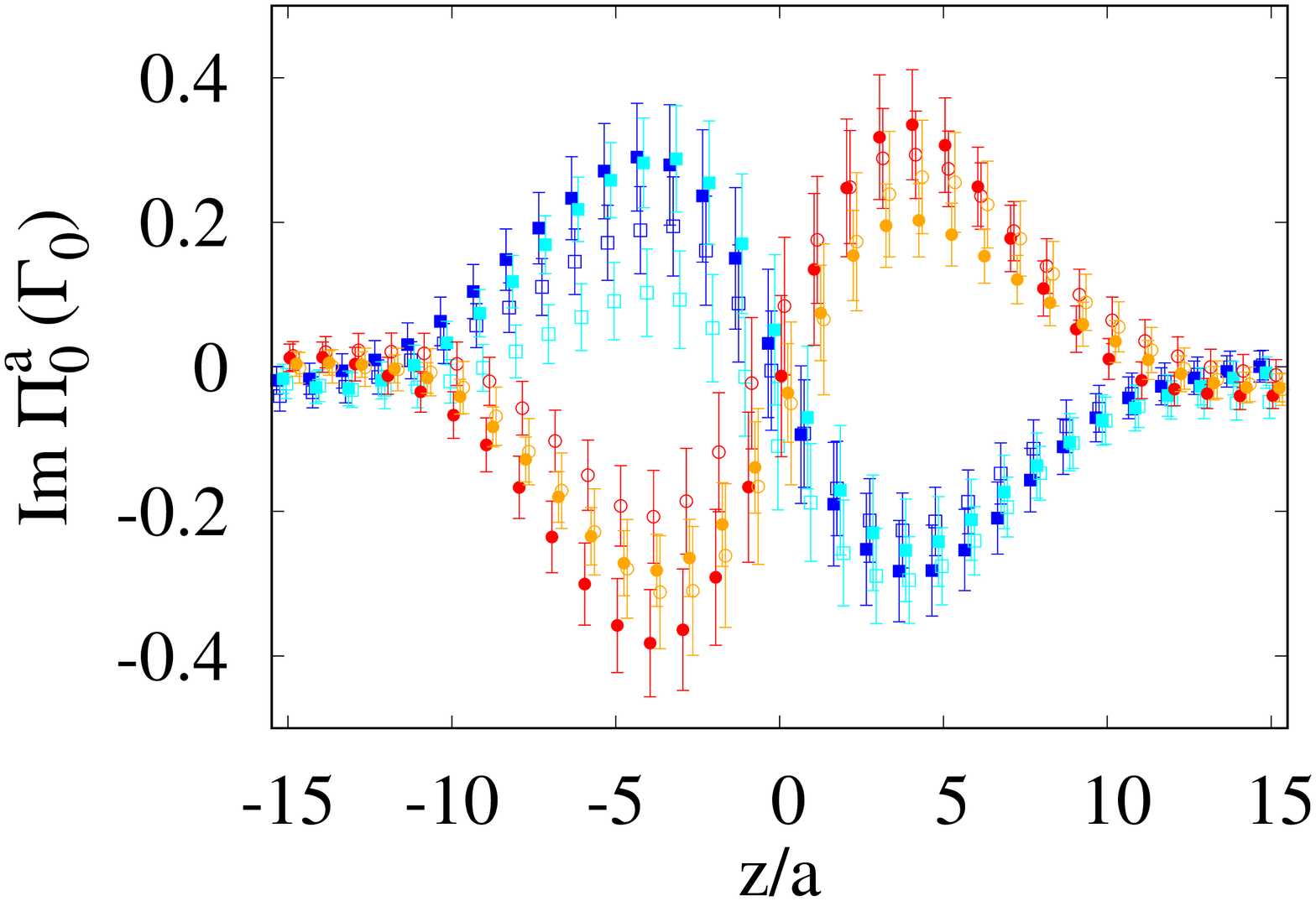}
\caption{Real (top) and imaginary (bottom) part of MEs $\Pi_0(\Gamma_0)$ in the symmetric (left) and asymmetric (right) frame. The nucleon boost is $|P_3|=1.25$ GeV and the momentum transfer is $-t=0.69$ GeV$^2$ (symmetric) or $-t=0.64$ GeV$^2$ (asymmetric). Different colors correspond to different permutations of $\vec{\Delta}$, i.e.\ $(2,0,0)$, $(-2,0,0)$, $(0,2,0)$ and $(0,-2,0)$ (in units of $2\pi/L$) at $P_3=|P_3|$ (bluish) or $P_3=-|P_3|$ (reddish).}
\label{fig:g0unpol}
\end{figure}

\begin{figure}[t!]
\centering
\includegraphics[angle=270,width=6.8cm]{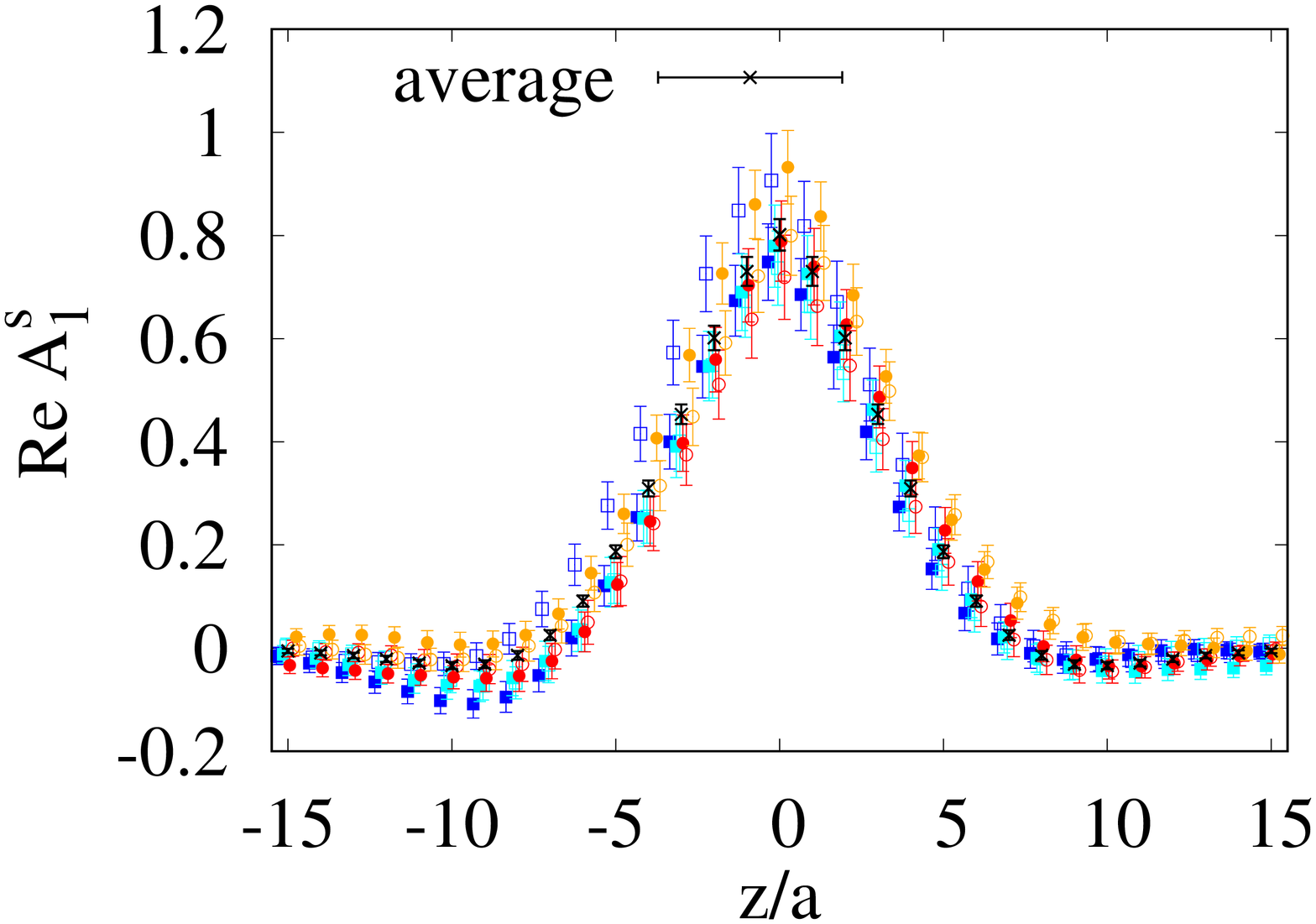}\hspace*{-5mm}
\includegraphics[angle=270,width=6.8cm]{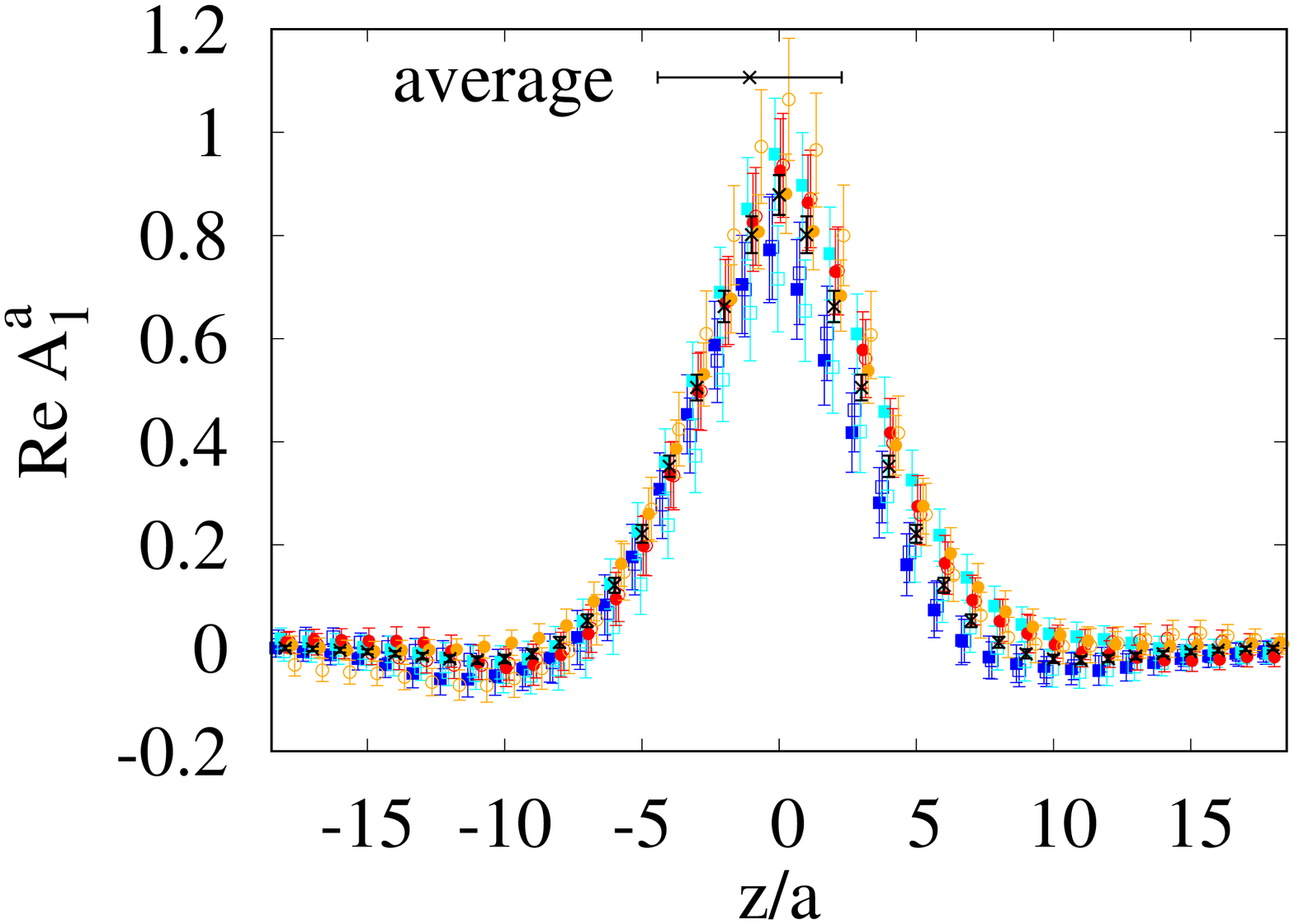}\vspace*{-3mm}
\includegraphics[angle=270,width=6.8cm]{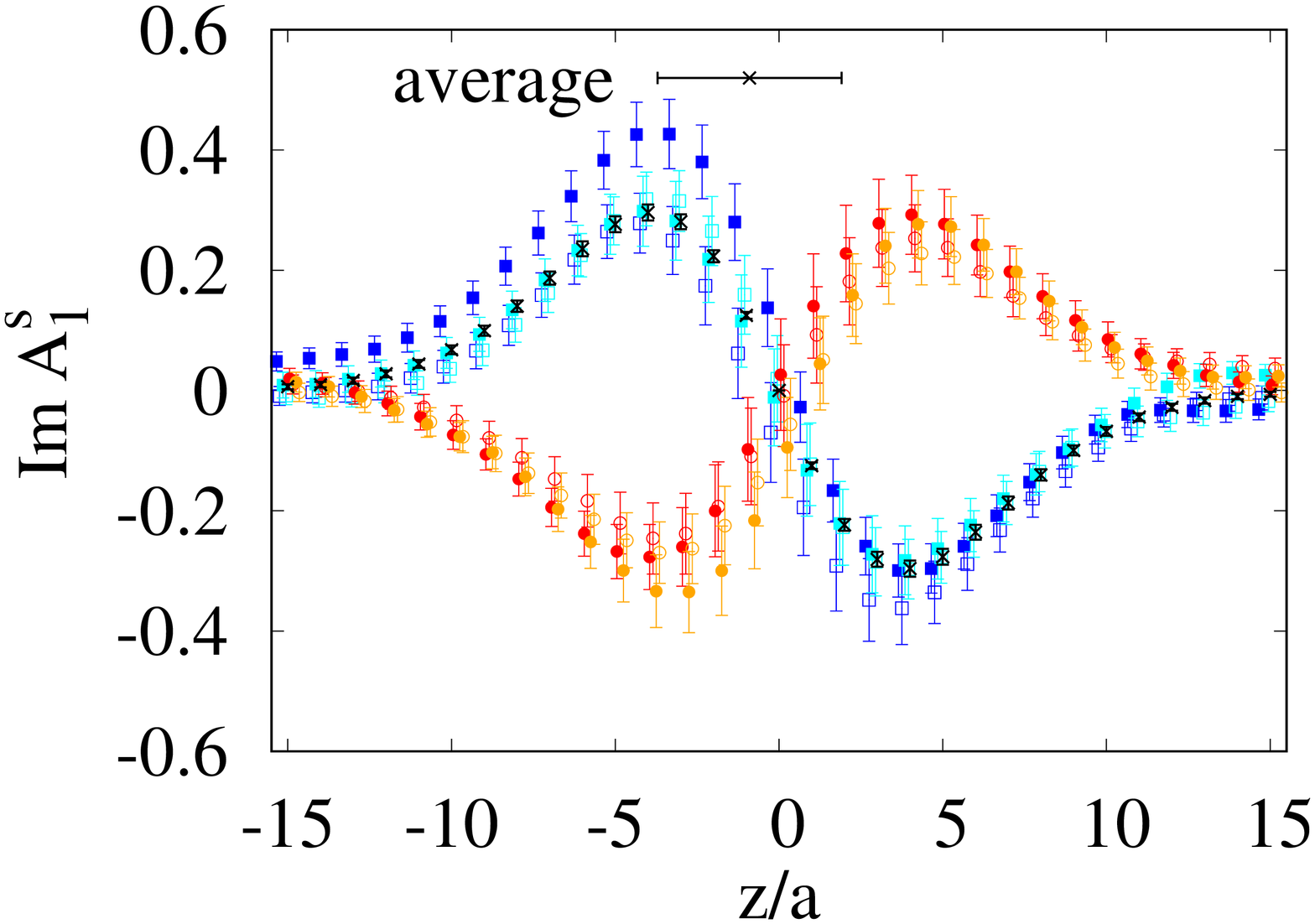}\hspace*{-5mm}
\includegraphics[angle=270,width=6.8cm]{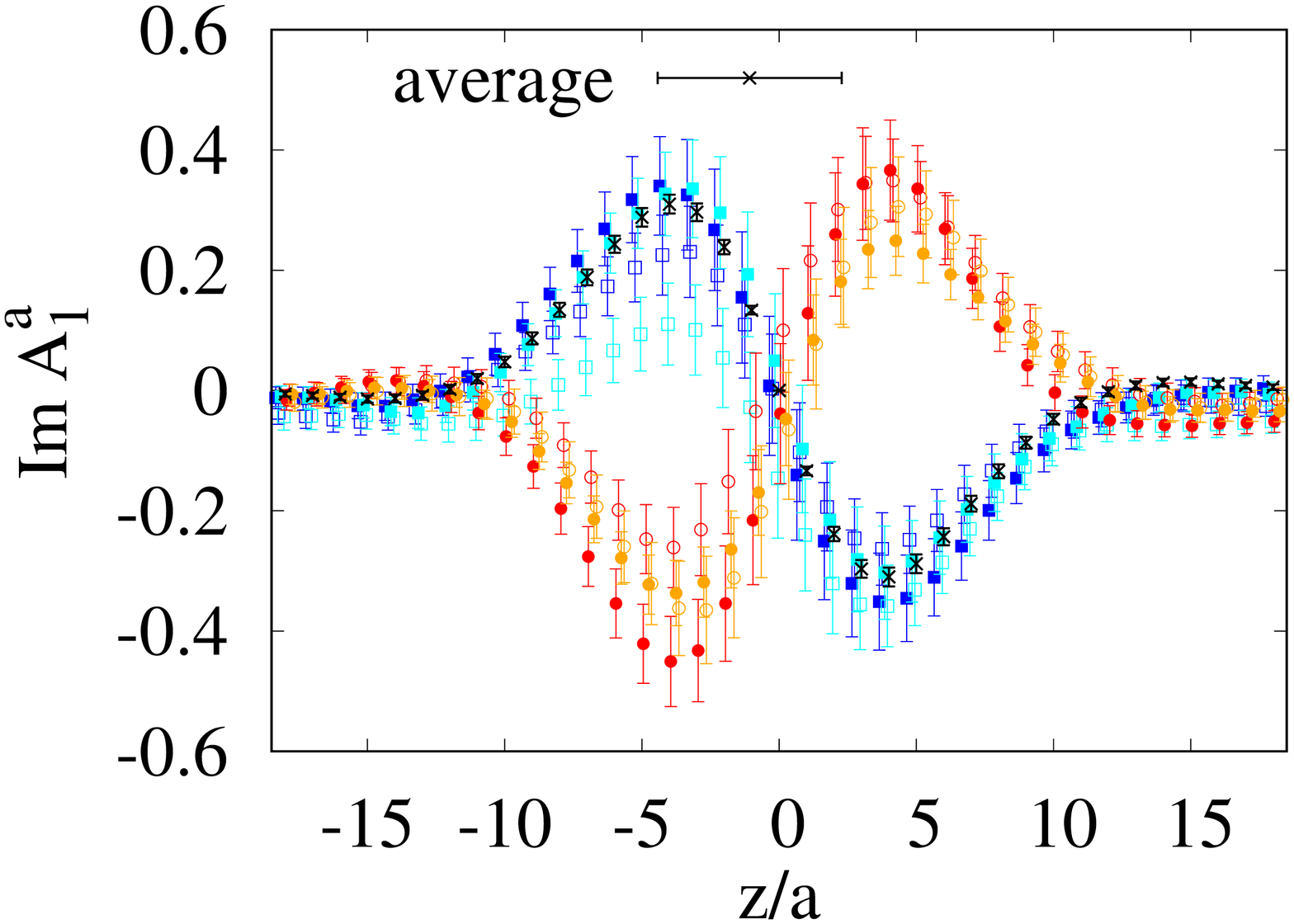}
\caption{Real (top) and imaginary (bottom) part of the amplitude $A_1$ extracted in the symmetric (left) and asymmetric (right) frame. The parameters and color conventions in the plot are the same as in Fig.~\ref{fig:g0unpol}; additionally shown are amplitudes averaged over the 8 kinematic setups, denoted with black points.}
\label{fig:A1}
\end{figure}

\begin{figure}[t!]
\centering
\includegraphics[angle=270,width=6.8cm]{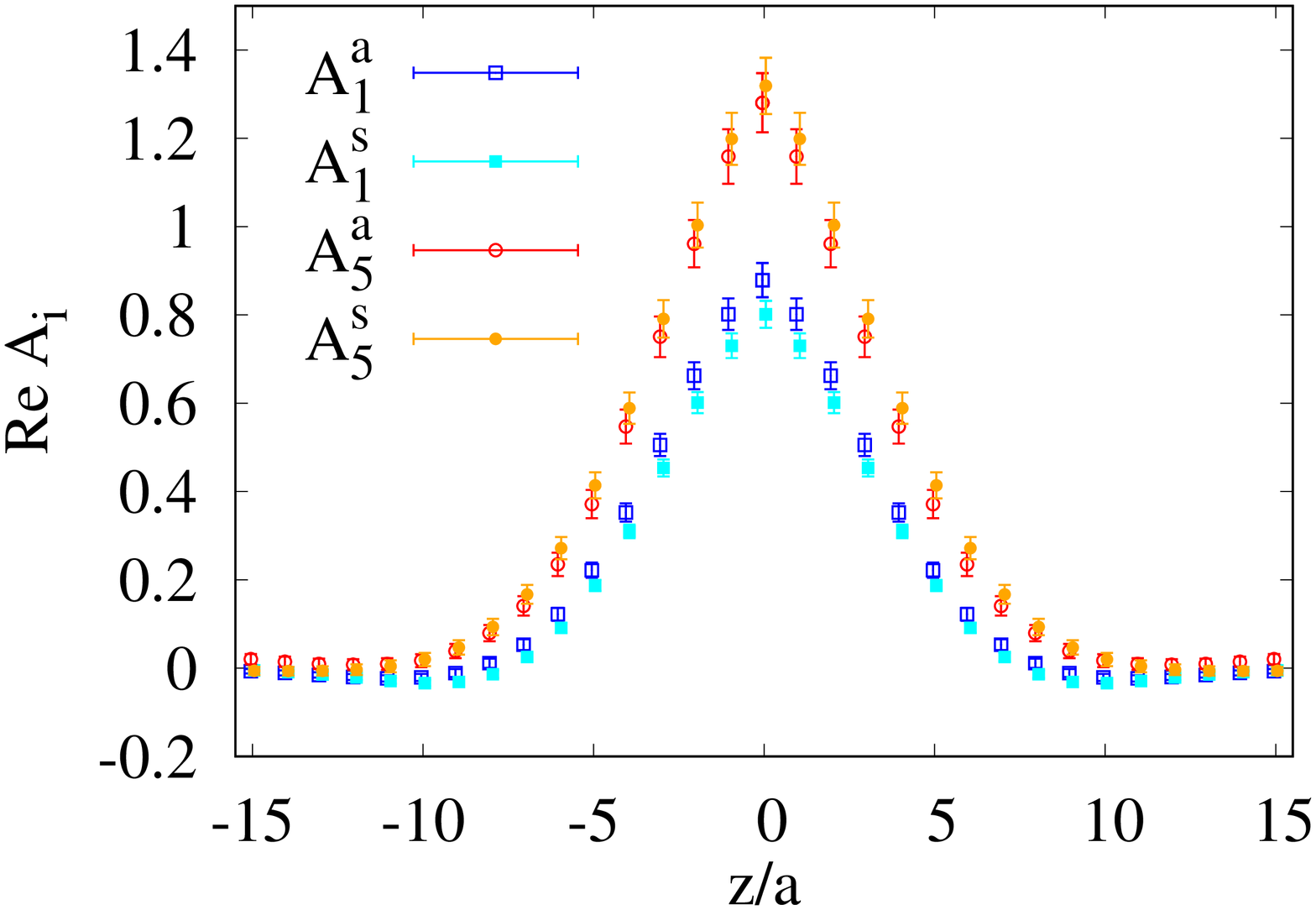}\hspace*{-5mm}
\includegraphics[angle=270,width=6.8cm]{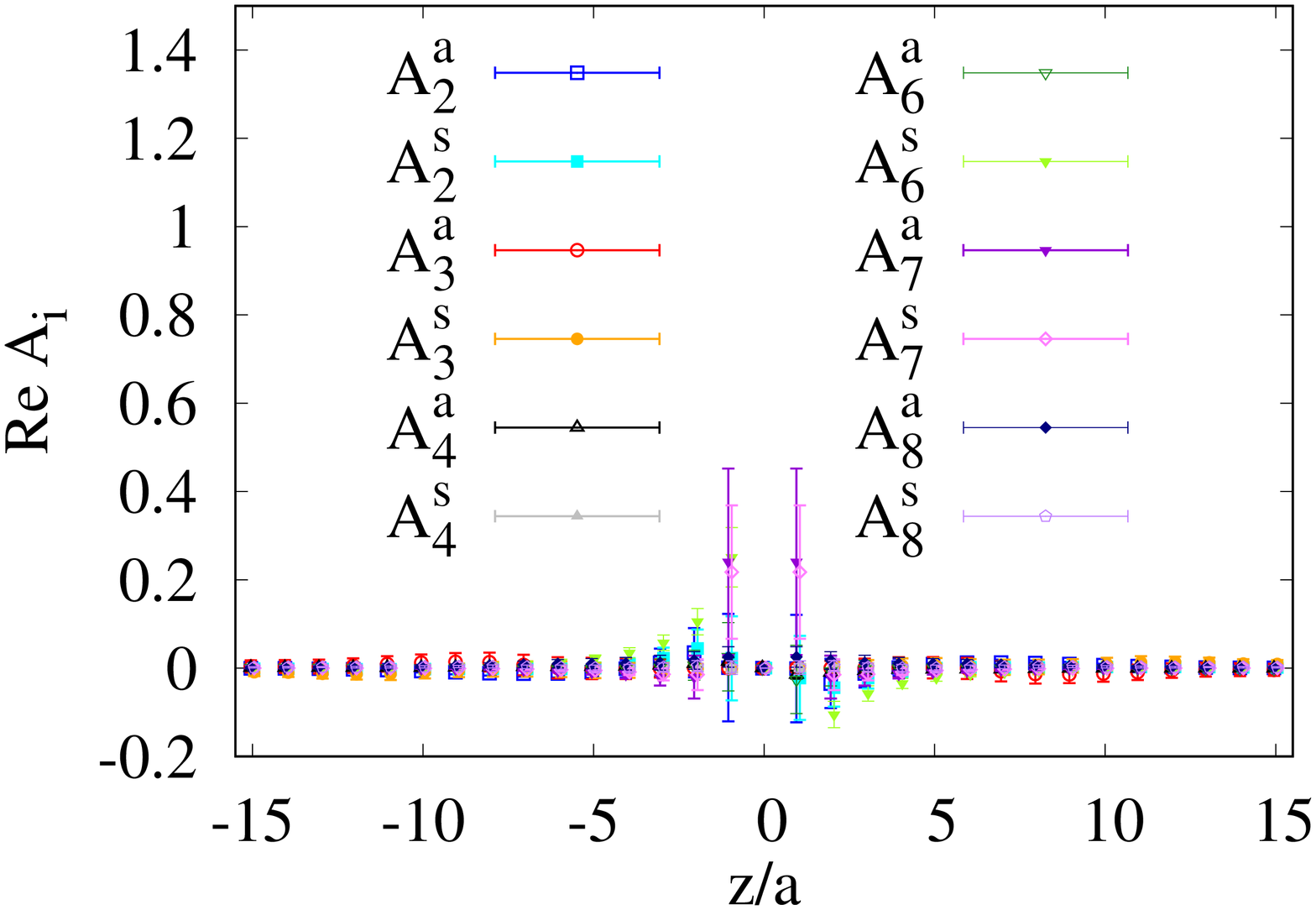}\vspace*{-3mm}
\includegraphics[angle=270,width=6.8cm]{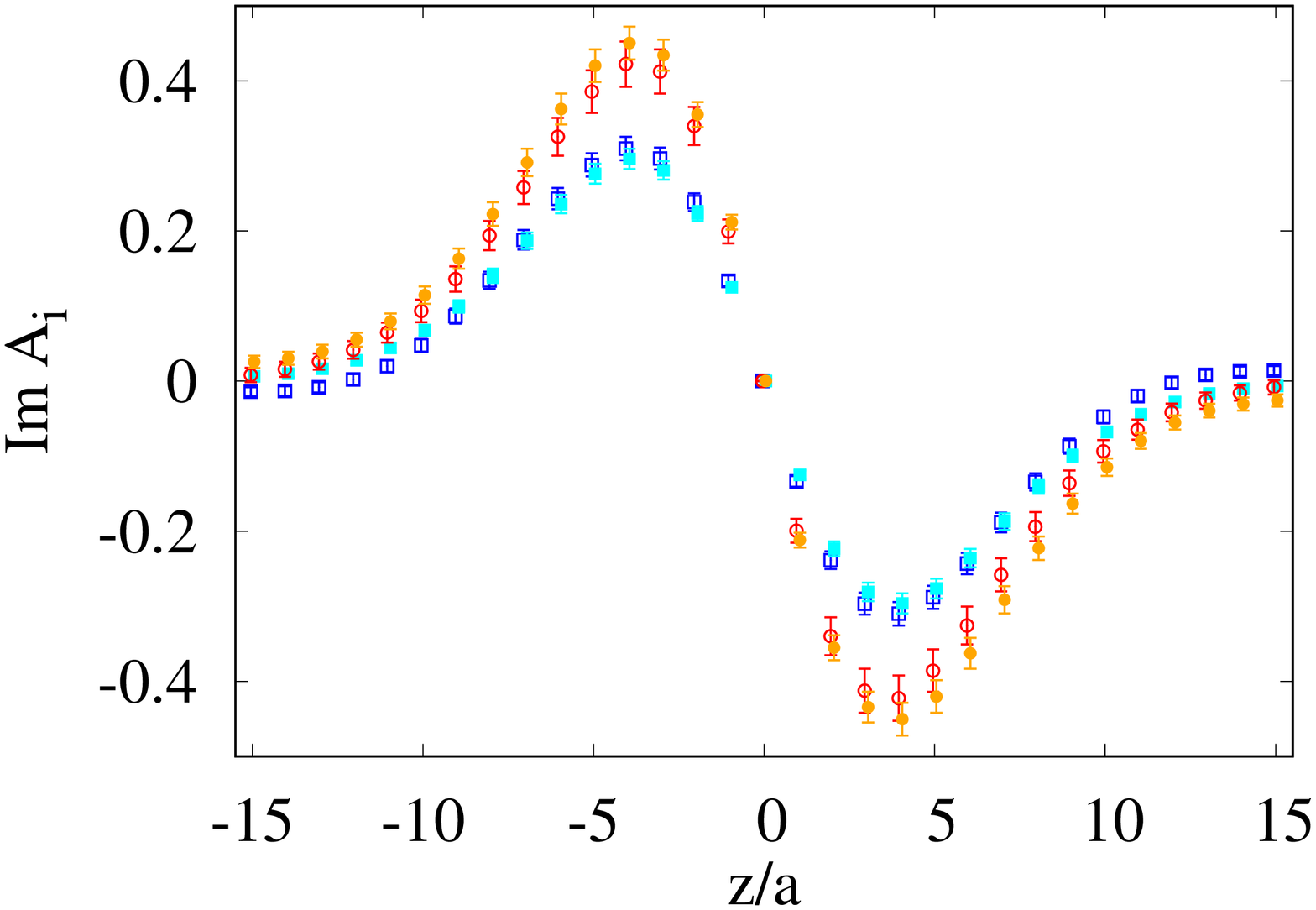}\hspace*{-5mm}
\includegraphics[angle=270,width=6.8cm]{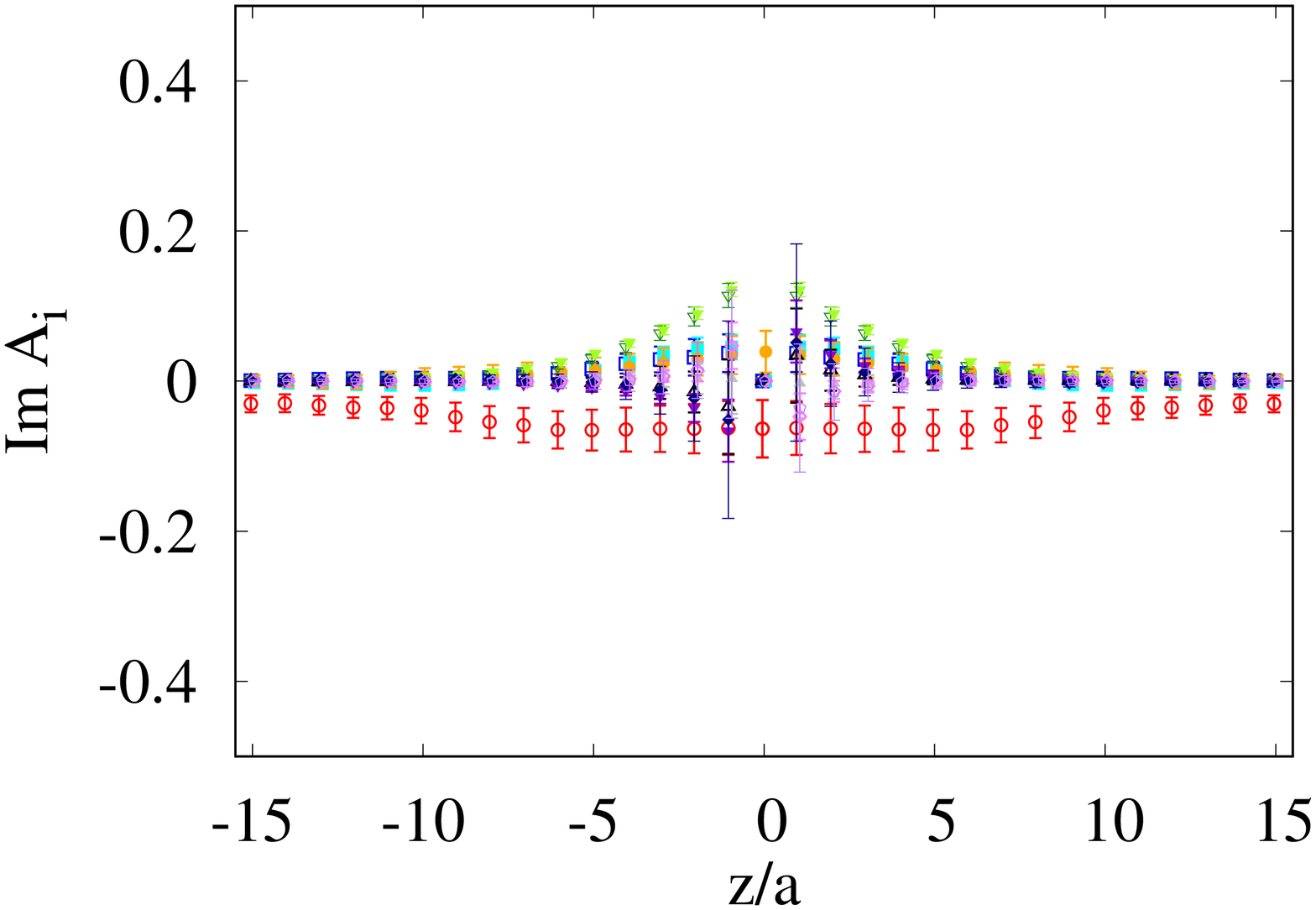}
\caption{Real (top) and imaginary (bottom) part of the amplitudes $A_i$ compared between the asymmetric (darker colors) and symmetric frame (lighter colors). The left part shows the amplitudes with largest values, $A_1$ and $A_5$, and the right part the remaining ones, $A_2,A_3,A_4,A_6,A_7,A_8$, all averaged over the 8 kinematic setups. The nucleon boost is $|P_3|=1.25$ GeV and the momentum transfer is $-t=0.69$ GeV$^2$ (symmetric) or $-t=0.64$ GeV$^2$ (asymmetric).}
\label{fig:As}
\end{figure}

\begin{figure}[t!]
\centering
\includegraphics[angle=270,width=6.8cm]{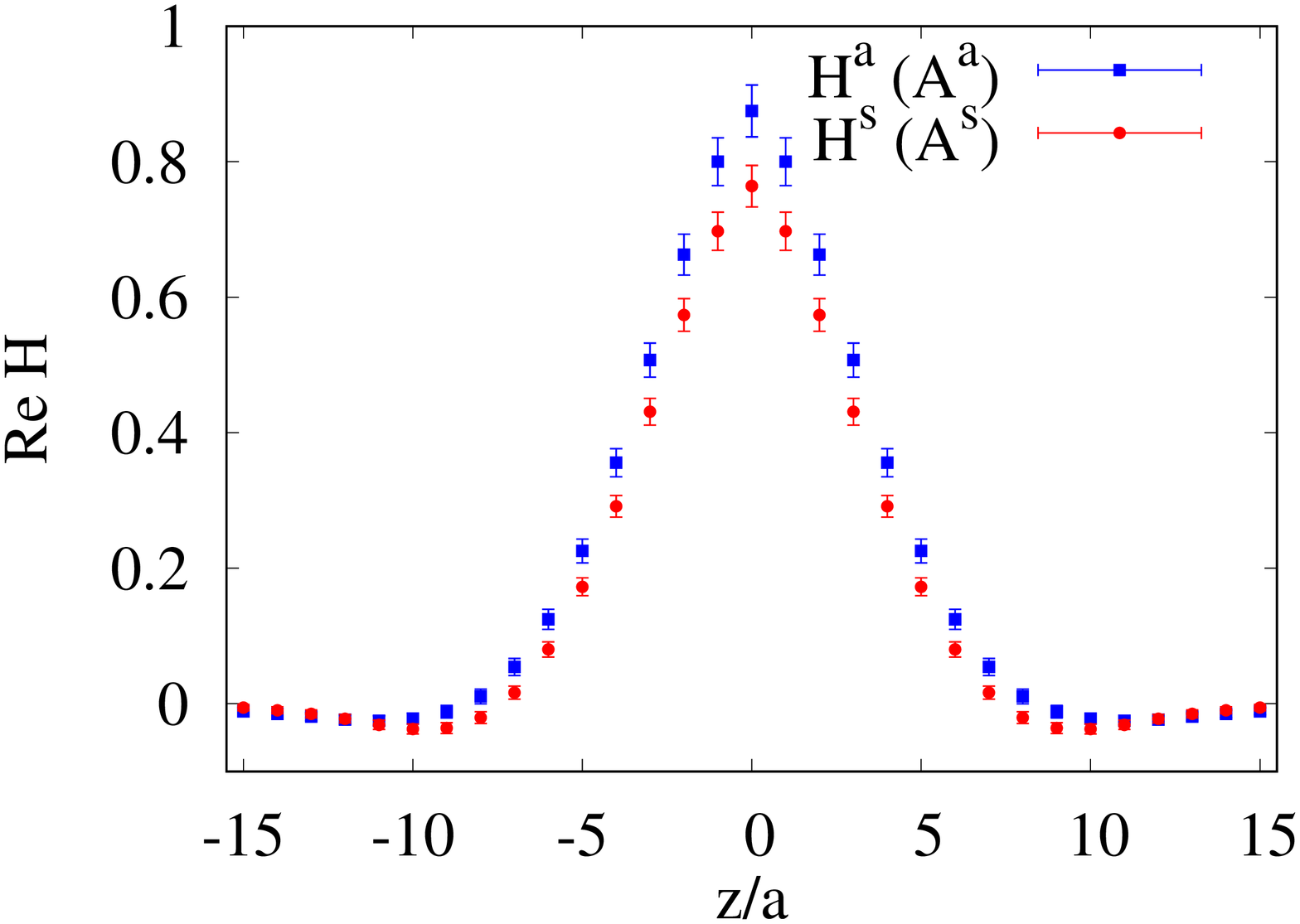}\hspace*{-5mm}
\includegraphics[angle=270,width=6.8cm]{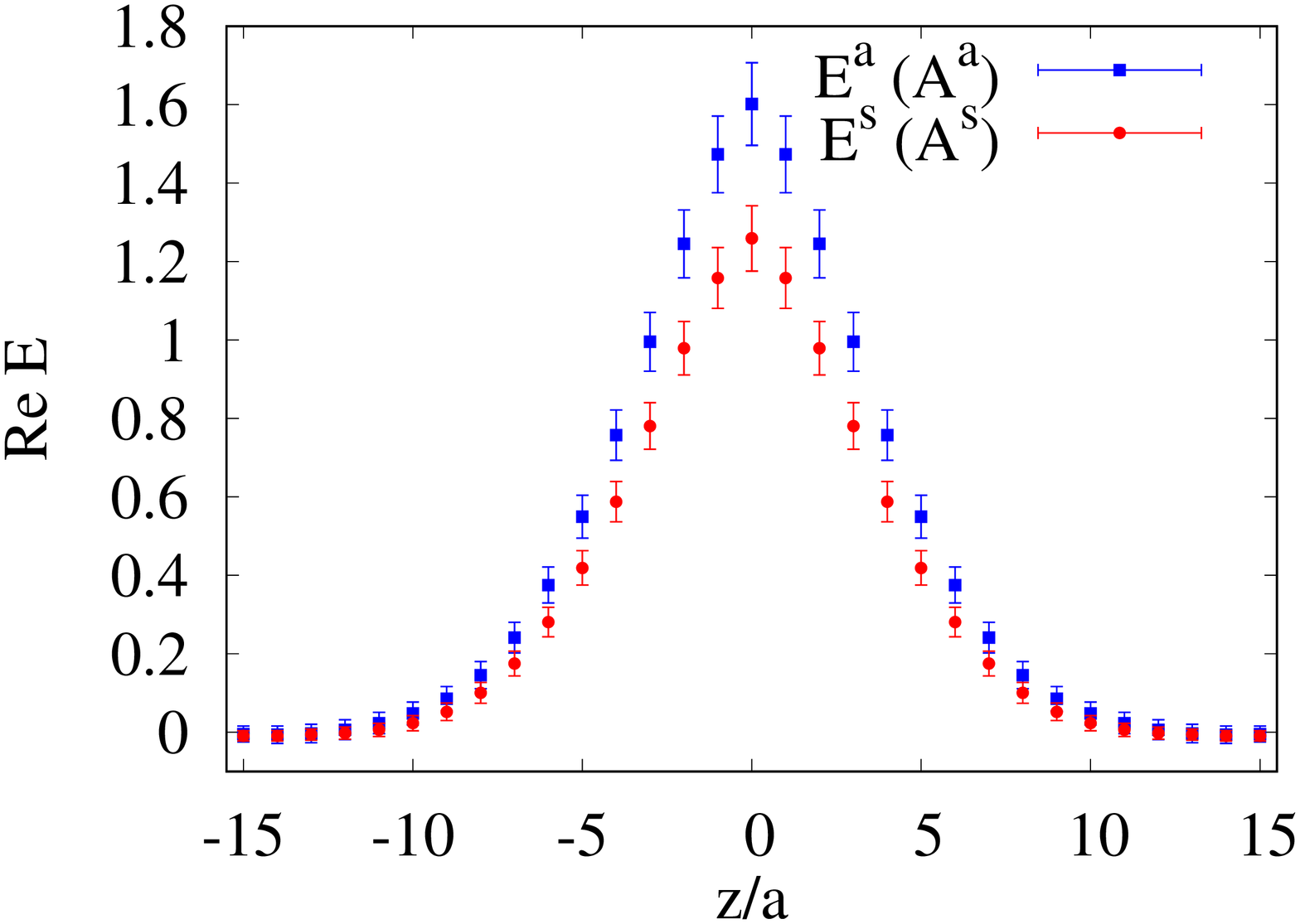}\vspace*{-3mm}
\includegraphics[angle=270,width=6.8cm]{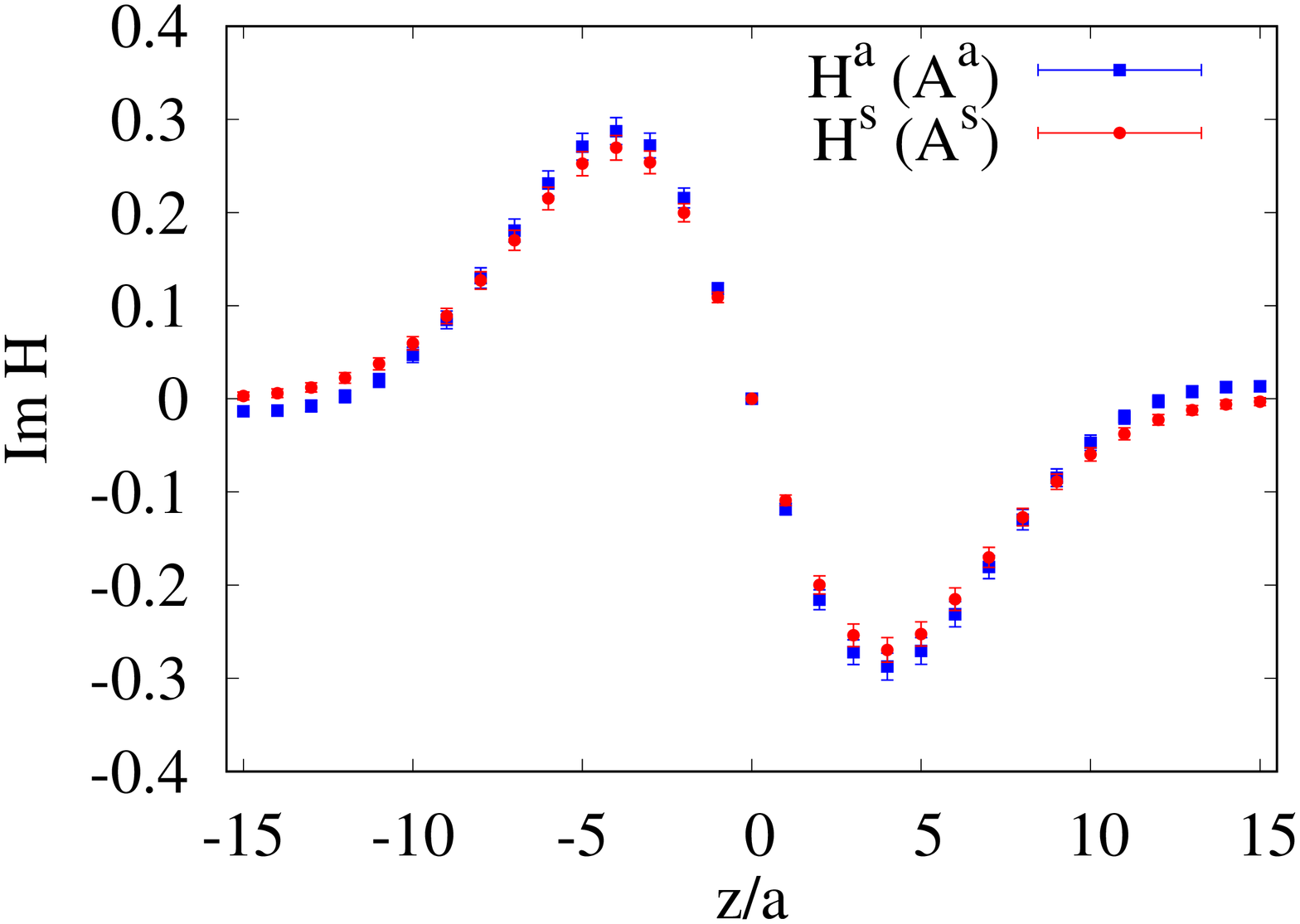}\hspace*{-5mm}
\includegraphics[angle=270,width=6.8cm]{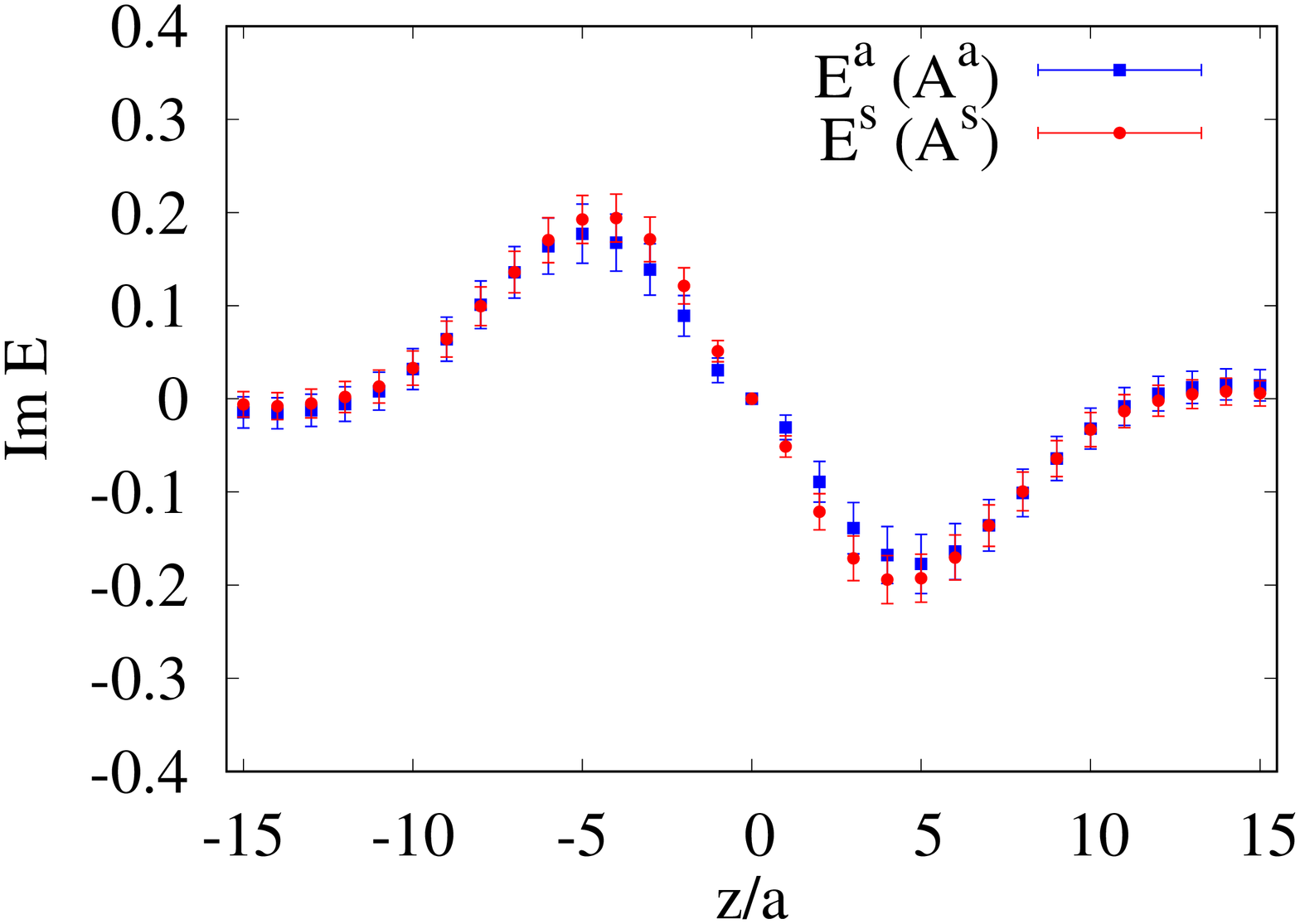}
\caption{Comparison of real (top) and imaginary (bottom) parts of the $H$ (left) and $E$ (right) quasi-GPDs in coordinate space. The standard definition of $H/E$ is used. The nucleon boost is $|P_3|=1.25$ GeV and the momentum transfer is $-t=0.69$ GeV$^2$ (symmetric) or $-t=0.64$ GeV$^2$ (asymmetric).}
\label{fig:HEstd}
\end{figure}

\begin{figure}[t!]
\centering
\includegraphics[angle=270,width=6.8cm]{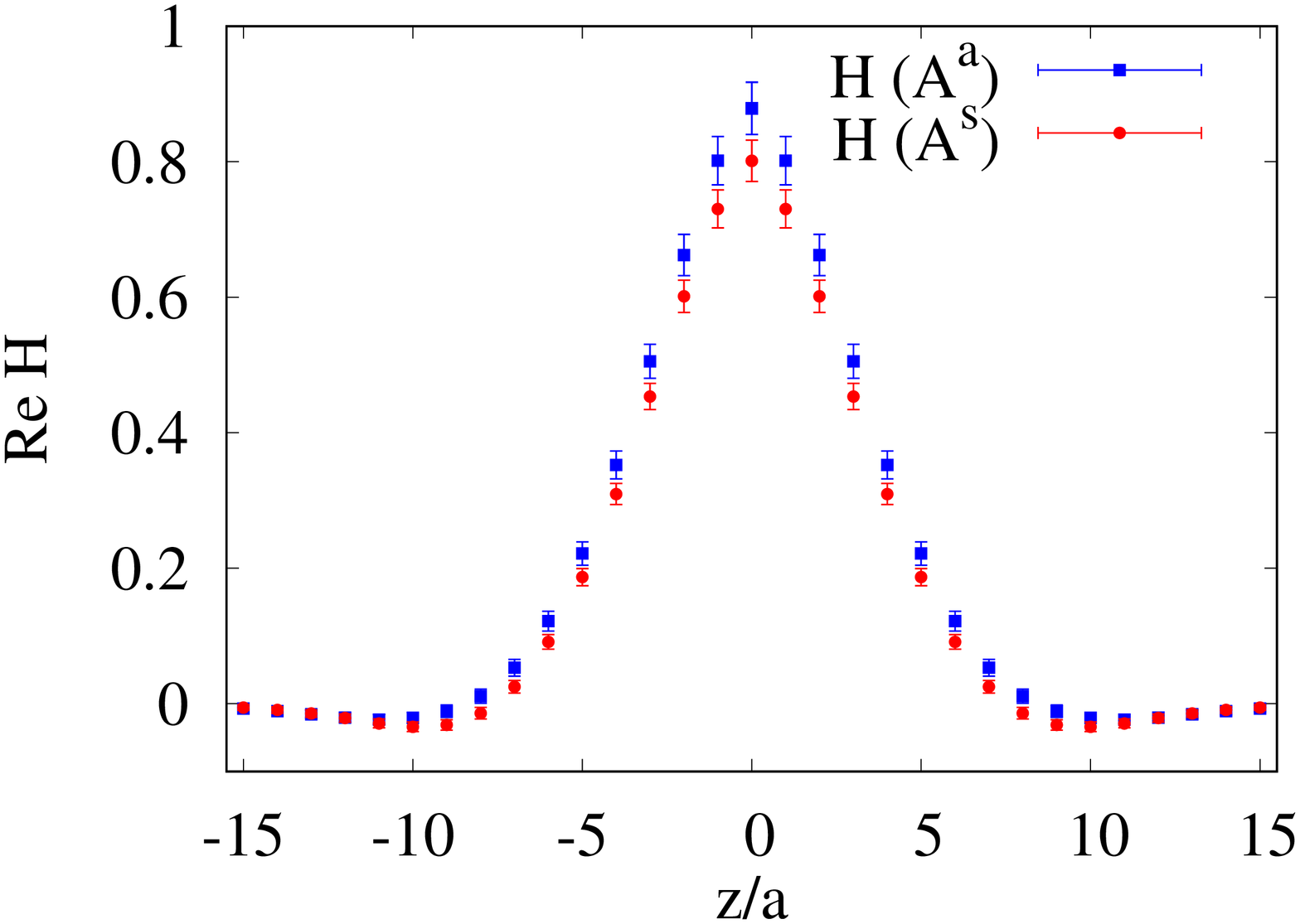}\hspace*{-5mm}
\includegraphics[angle=270,width=6.8cm]{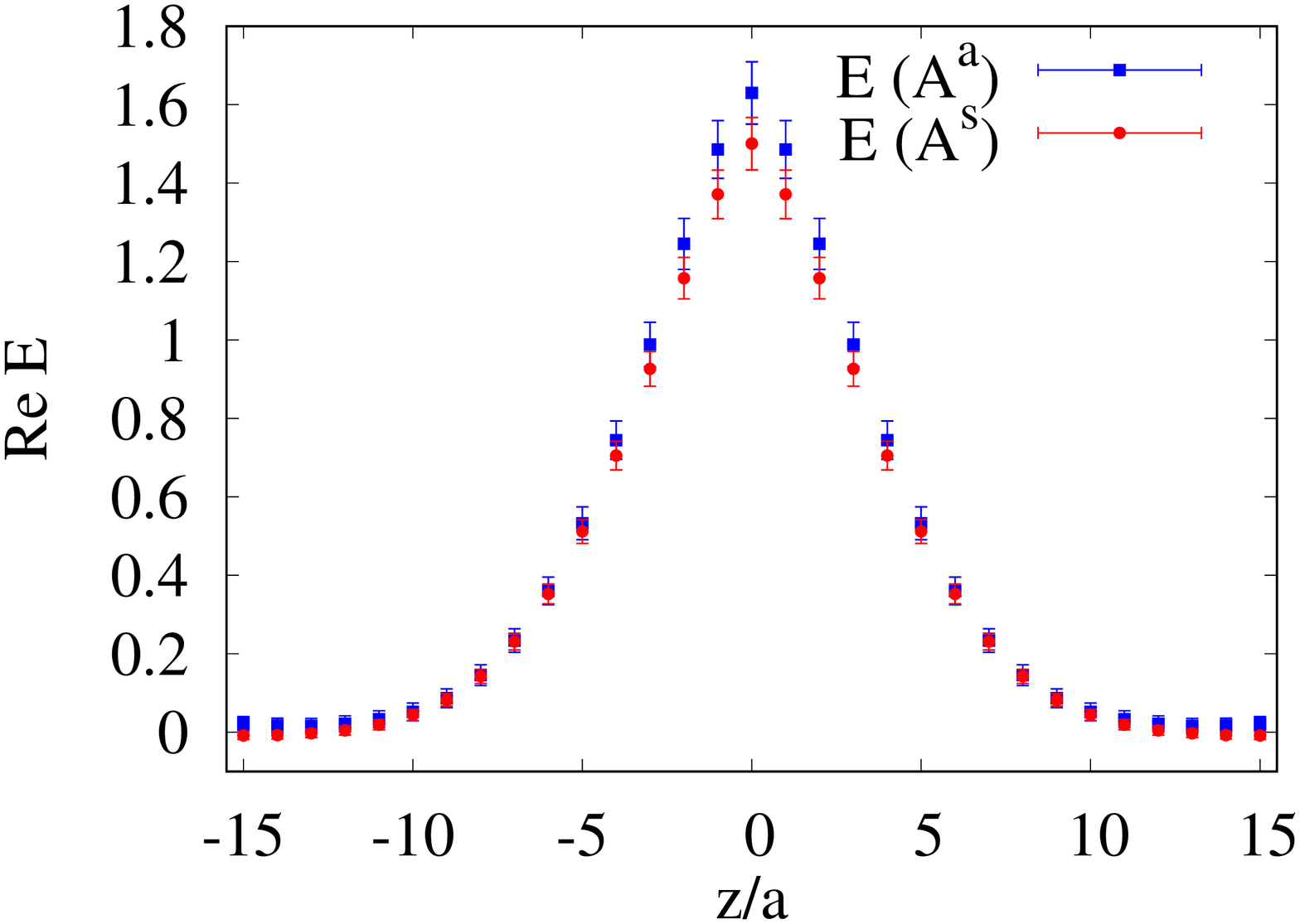}\vspace*{-3mm}
\includegraphics[angle=270,width=6.8cm]{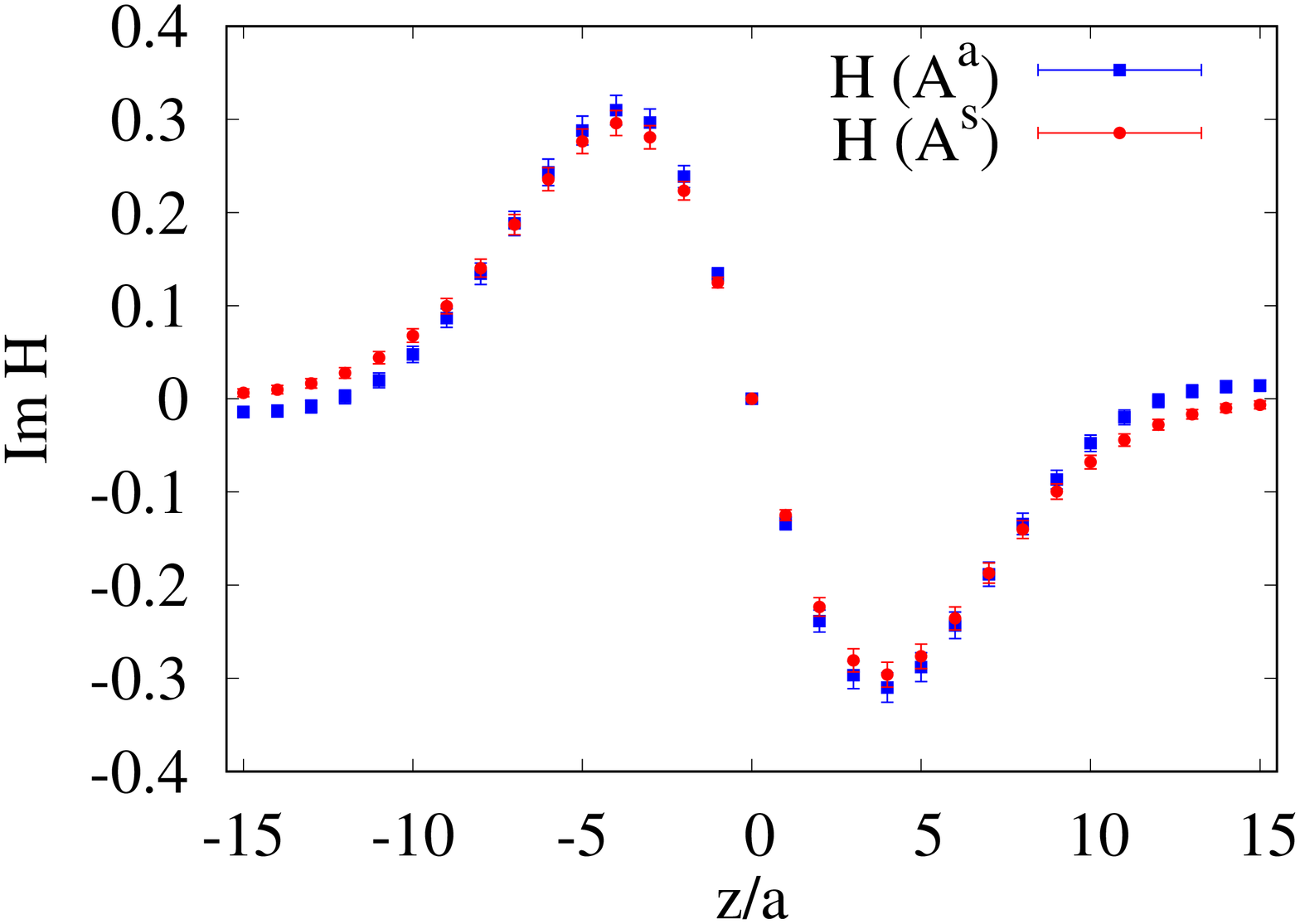}\hspace*{-5mm}
\includegraphics[angle=270,width=6.8cm]{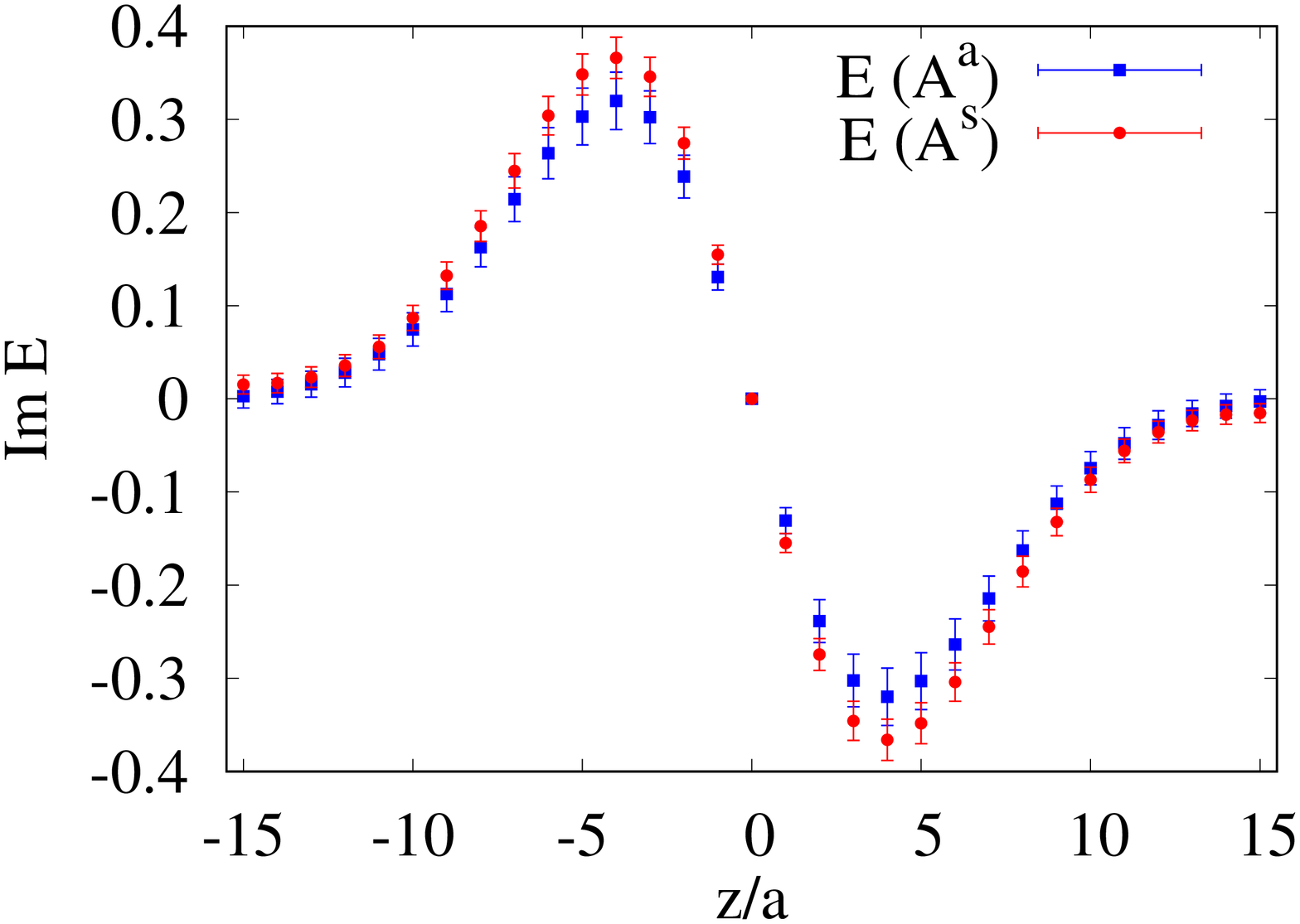}
\caption{Comparison of real (top) and imaginary (bottom) parts of the $H$ (left) and $E$ (right) quasi-GPDs in coordinate space. The Lorentz-invariant definition of $H/E$ is used. The nucleon boost is $|P_3|=1.25$ GeV and the momentum transfer is $-t=0.69$ GeV$^2$ (symmetric) or $-t=0.64$ GeV$^2$ (asymmetric).}
\label{fig:HELI}
\end{figure}

\begin{figure}[t!]
\centering
\includegraphics[angle=0,width=10cm]{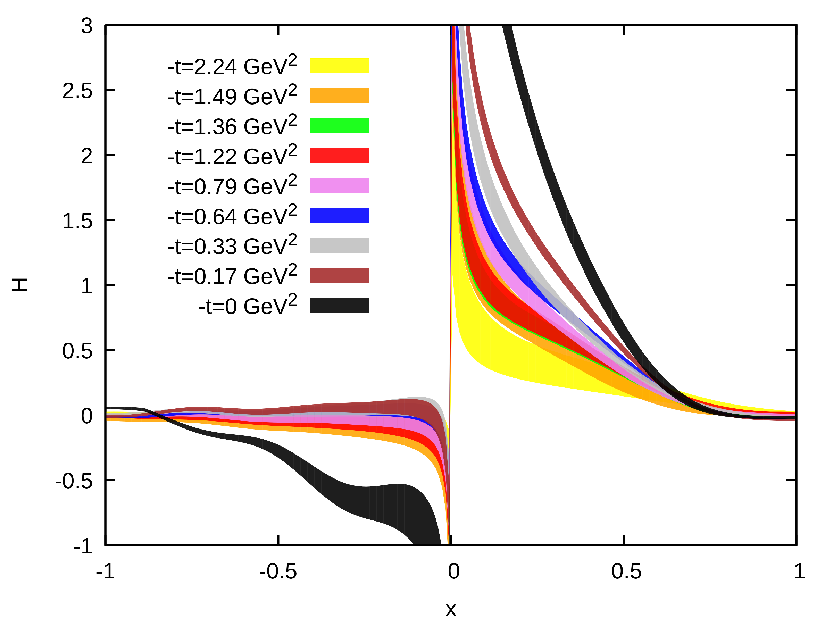}\vspace*{-5mm}\\
\includegraphics[angle=0,width=10cm]{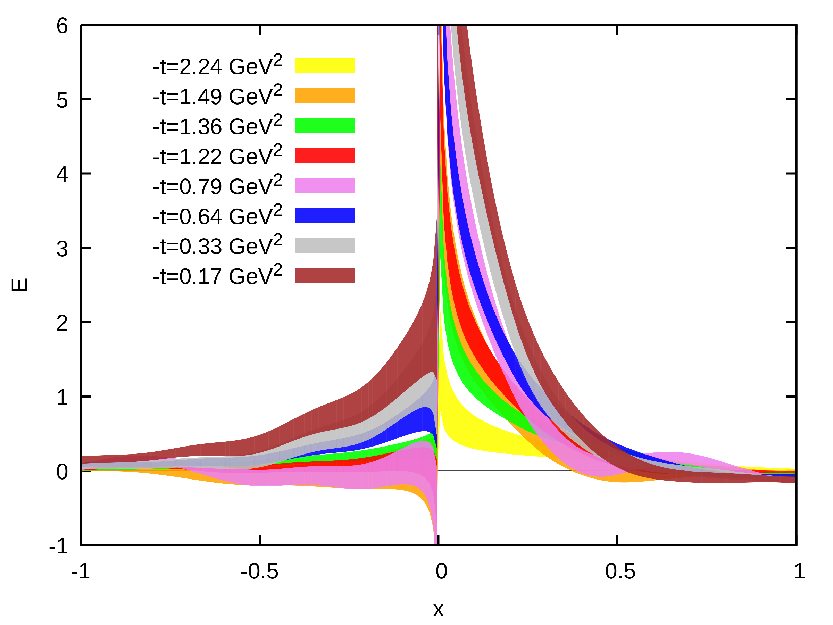}
\caption{Light-cone GPDs $H$ (top) and $E$ (bottom) for different momentum transfers $-t$. Data for the asymmetric frame at the nucleon boost $|P_3|=1.25$ GeV and the Lorentz-invariant definition. For the $H$ function, we also show the corresponding PDF ($-t=0$).}
\label{fig:tdep}
\end{figure}

\begin{figure}[t!]
\centering
\includegraphics[angle=270,width=6.8cm]{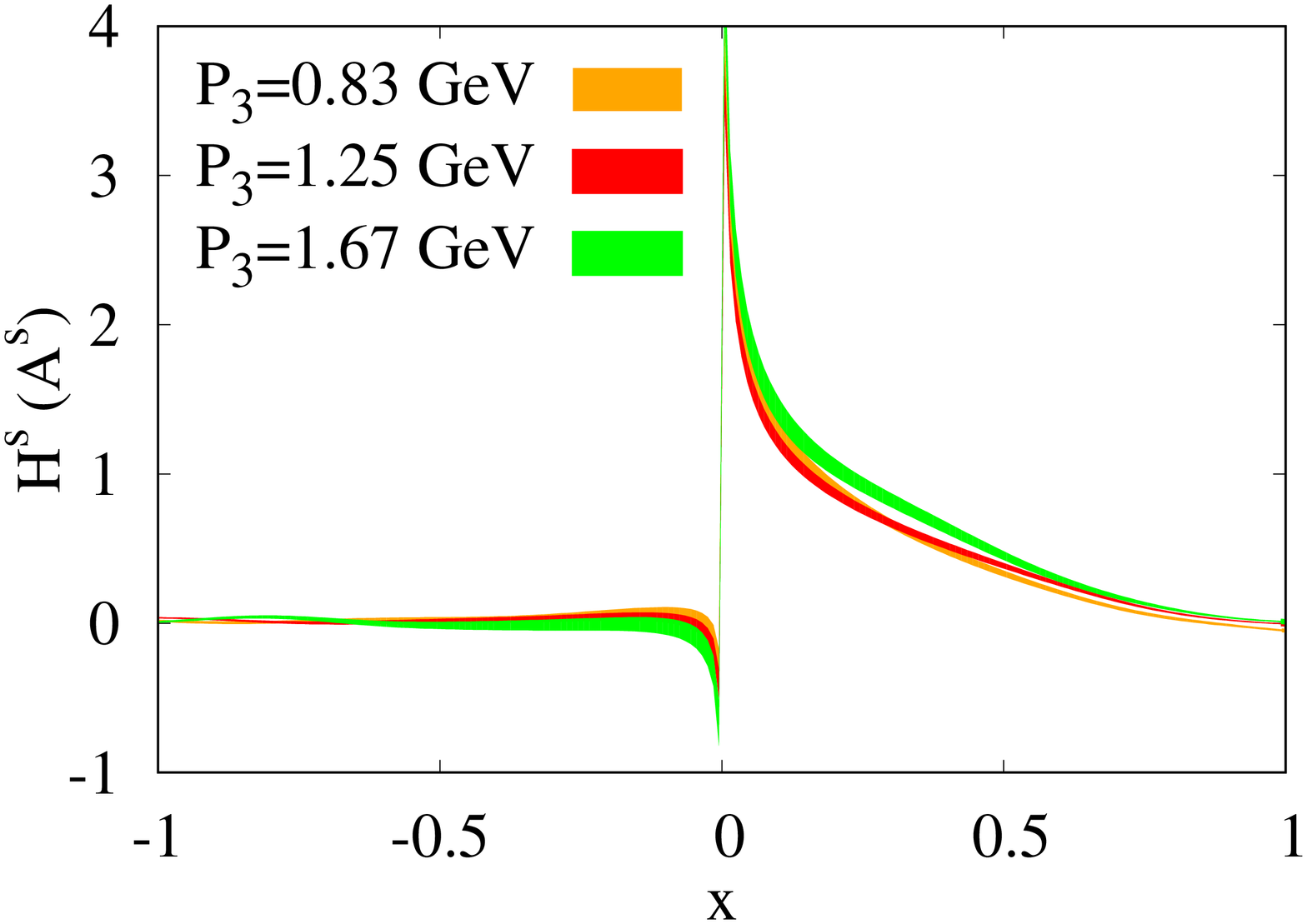}\hspace*{-5mm}
\includegraphics[angle=270,width=6.8cm]{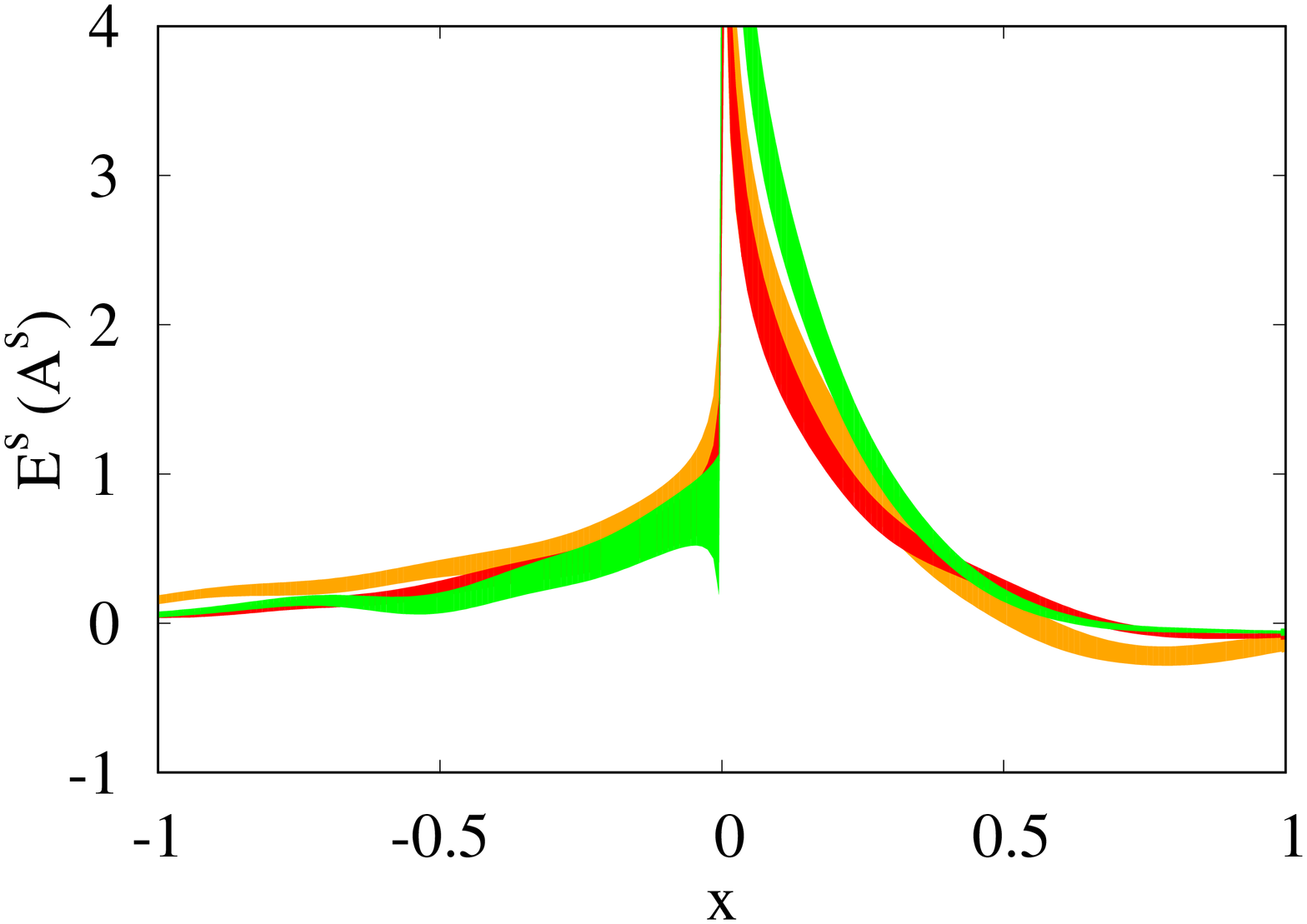}\vspace*{-3mm}
\includegraphics[angle=270,width=6.8cm]{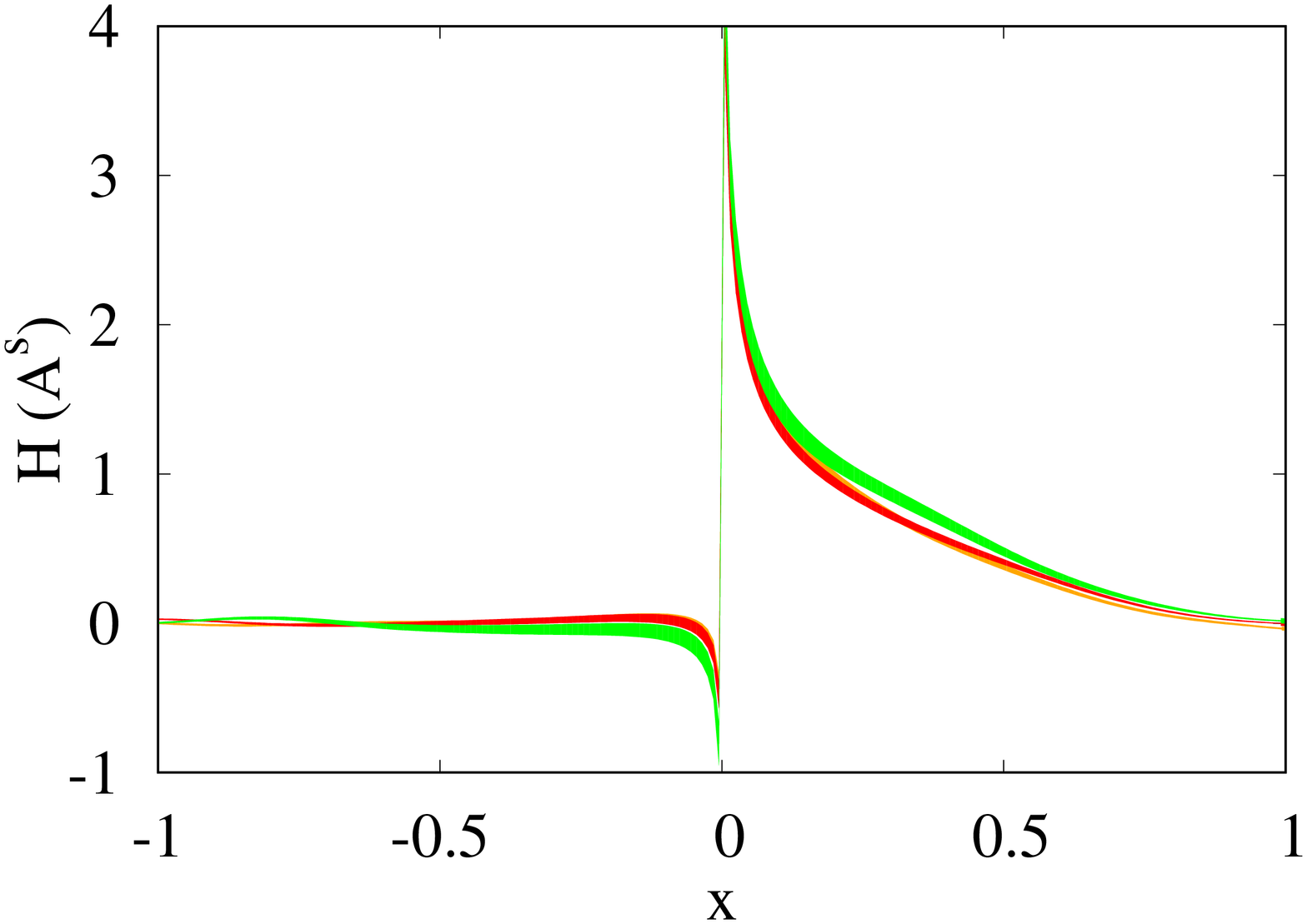}\hspace*{-5mm}
\includegraphics[angle=270,width=6.8cm]{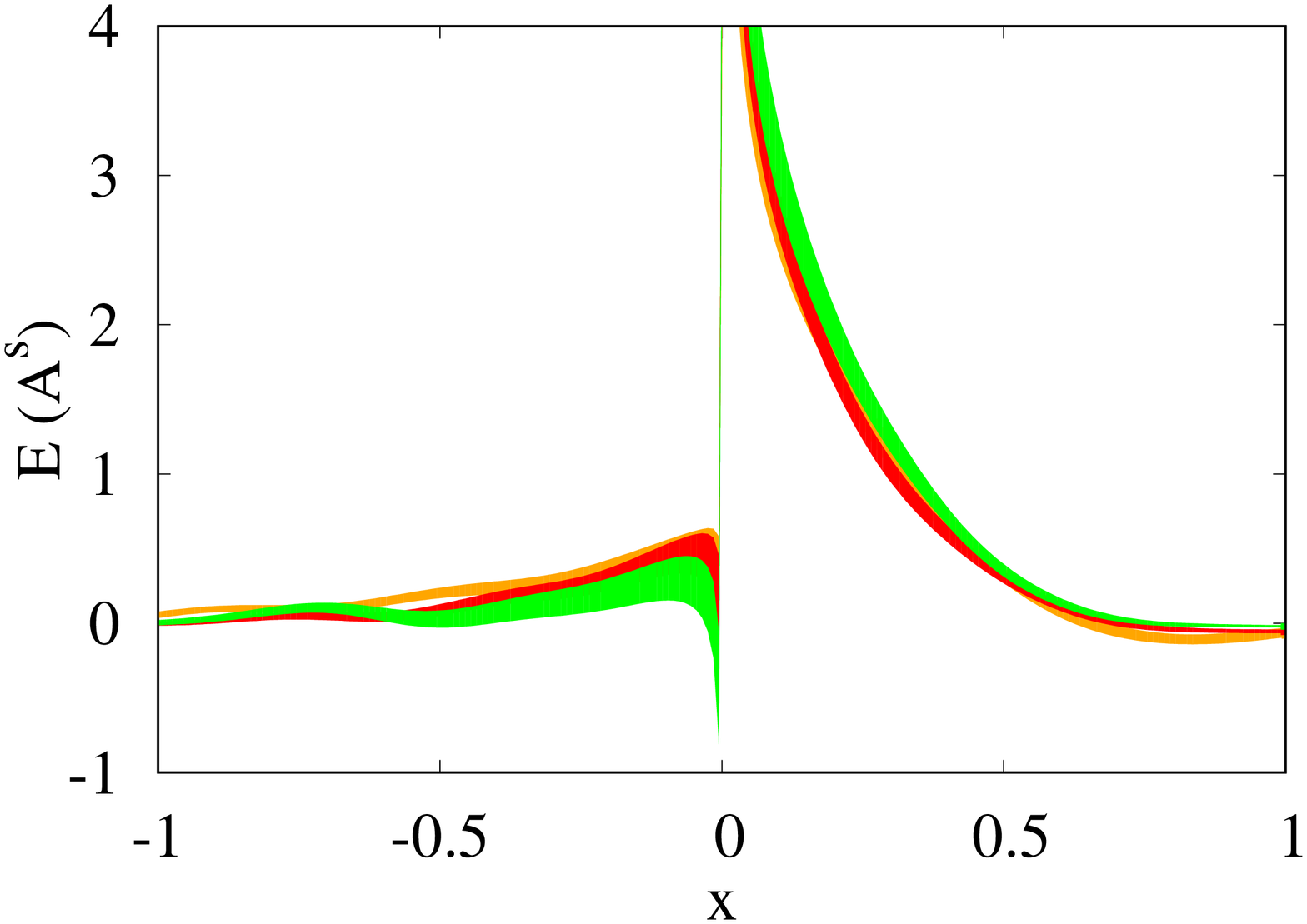}
\caption{Light-cone GPDs $H$ (left) and $E$ (right) from the standard definition (top) and the Lorentz-invariant definition (bottom). Data for the symmetric frame at three nucleon boosts $|P_3|=0.83,1.25,1.67$ GeV.}
\label{fig:convergence}
\end{figure}

We start by showing an example of bare MEs for the $\gamma_0$ insertion and the unpolarized projector, see Fig.~\ref{fig:g0unpol}.
The data are obtained in 8 kinematic setups, corresponding to 4 different permutations of $\vec{\Delta}$ and 2 directions of $P_3$.
As shown in Eqs.~(\ref{eq:P0G0sym}), (\ref{eq:P0G0asym}), the MEs in different frames are not equivalent, containing contributions from different amplitudes.
Moreover, the data in the symmetric frame have definite symmetries upon $P_3z\rightarrow -P_3z$, while in the asymmetric frame, these symmetry properties are lost.

In Fig.~\ref{fig:A1}, we show an analogous plot for the amplitude $A_1$, as extracted from both frames.
At this level, the definite symmetry properties are recovered and the comparison between $A_i$'s extracted from both frames becomes meaningful, as the amplitudes are Lorentz-invariant.

A systematic comparison of all amplitudes is shown in Fig.~\ref{fig:As}.
According to theoretical expectations, the amplitudes are consistent upon extraction from both frames.
Note that, strictly speaking, the data from the symmetric and the asymmetric frame correspond to slightly different values of the momentum transfer ($-t=0.69$ GeV$^2$ vs.\ $-t=0.64$ GeV$^2$), due to the non-zero temporal component of $\Delta_\mu$ in the asymmetric frame ($E_i\neq E_f$).
However, within our statistical precision, these differences are not visible.

In Fig.~\ref{fig:HEstd}, we compare the $H$ and $E$ coordinate-space quasi-GPDs obtained according to the frame-dependent definitions (\ref{eq:Hsym})-(\ref{eq:Easym}).
As expected, the results from different frames do not need to agree at finite $P_3$ and the difference is clearly visible in the real part of both functions, while the effects in the imaginary part are smaller than our statistical precision.
Performing such a comparison for the Lorentz-invariant definitions (\ref{eq:HLI})-(\ref{eq:ELI}), see Fig.~\ref{fig:HELI}, the expected agreement between the frames is observed.

All the above results confirm numerically the validity of the method of Lorentz-invariant amplitudes and this proof-of-concept analysis can be viewed as a prelude to actual applications to extracting GPDs in a wide kinematic range.
In the remainder of this section, we show preliminary results for two follow-up aspects.

The first one corresponds to the original motivation of making the GPD calculations more efficient in terms of computer time, by accessing several values of the momentum transfer in a single determination.
In Fig.~\ref{fig:tdep}, we show the $t$-dependence of the $H$ and $E$ light-cone GPDs in momentum space, i.e.\ after applying all 5 steps of the full procedure outlined in Section \ref{sec:procedure} (so far, we concentrated on steps 1 and 2, leading to coordinate-space quasi-GPDs).
We emphasize that all curves in Fig.~\ref{fig:tdep} were obtained with a single sink momentum of $\vec{P}=(0,0,3)(2\pi/L)$ in the asymmetric frame.
In the plot for the GPD $H$, we also include its forward limit, i.e.\ the unpolarized PDF.
In the positive-$x$ region (quarks), we note that both GPDs are quickly suppressed with an increasing $-t$ value for small-to-intermediate $x$.
The GPD $E$ decays overall more quickly to zero with increasing $x$, with its values vanishing around $x\approx0.5$ regardless of $-t$, whereas the GPD $H$ (including the PDF) vanishes only around $x\approx0.7$.
The negative-$x$ part (antiquarks at positive $x$) shows little dependence in the $H$ function, apart from a significantly different result in the forward limit.
The latter is attributed to a large systematic effect, also observed in our earlier PDF analyses \cite{Alexandrou:2018pbm,Alexandrou:2019lfo,Alexandrou:2020qtt} and attributed to a large extent to discretization effects.
Overall, the negative-$x$ region should be treated with caution and requires further analyses to better understand the systematic effects.
Hence, we also treat the $x<0$ results for the GPD $E$ with caution -- the present results seem to suggest a clearly non-zero value of $E$ at small negative $x$ and small momentum transfers, the robustness of this result needs to be confirmed.
The $t$-dependence that we obtain, upon Fourier transform, will be translated to the spatial tomography of quarks in the transverse plane.

Another aspect under scrutiny in the present work was the question about the convergence of the standard and Lorentz-invariant definitions of $H$ and $E$. 
This was investigated by comparing results from three nucleon boosts, $P_3=0.83,1.25,1.67$ GeV, in the symmetric frame at $-t=0.69$ GeV$^2$.
In the left panel of Fig.~\ref{fig:convergence}, this comparison is shown for the $H$ function, top/bottom panel for the standard/Lorentz-invariant definition.
We observe a very similar picture for both definitions, i.e.\ practically equivalent convergence properties.
Note, however, that the two lowest boosts are in perfect agreement with each other, while the largest boost leads to different values in the range $0.2\lesssim x\lesssim0.5$.
This is an indication of a possible problem with convergence towards the light cone and warrants additional analyses at increased nucleon boosts.
The situation is qualitatively different in the GPD $E$ (right panel of Fig.~\ref{fig:convergence}).
In this case, the reduced contribution from the amplitude $A_6$ (compare Eqs.~(\ref{eq:Esym}) and (\ref{eq:ELI})) has a significant impact on the final GPD, with the one utilizing the Lorentz-invariant definition evincing perfect agreement among all boosts, contrary to the standard definition that leads to different results at the analyzed $P_3$ values.
Moreover, there is significant impact of the used definition on the statistical quality of the signal.
This is attributed to the addition of the ME $\Pi_{1/2}(\Gamma_3)$ that cancels large part of the correlated noise in $\Pi_0(\Gamma_{1/2})$.
In the end, better convergence of the Lorentz-invariant definition is confirmed for the $E$ function, with little influence for the GPD $H$.
We emphasize that convergence properties of various definitions cannot be predicted \textit{a priori} -- the removal/reduction of contribution from certain amplitudes leads to a modification of higher-twist effects that may reduce these effects in the largest-value amplitudes ($A_1,A_5$ in the vector case), but may also be of no importance or even enhance them.

\section{Summary and prospects}
In these proceedings, we reported a new method of accessing GPDs from lattice QCD simulations.
It relies on the quasi-distribution method of Ji, but a novel aspect is to perform calculations in asymmetric frames of reference.
This offers a major advantage of obtaining several values of momentum transfer from a single calculation, by attributing the momentum transfer entirely to the source nucleon state.
We showed a proof-of-concept analysis with numerical evidence that the approach works according to expectations.
A byproduct of the method is the realization that alternative definitions of physical GPDs can be employed, differing in convergence properties to the light cone and as such, may converge better.

The directions for further analyses for unpolarized GPDs are multiple and include calculations for several momentum transfers and values of skewness, at different values of the nucleon boost, as well as investigation of other approaches to renormalization and matching (e.g.\ the hybrid scheme) and to the reconstruction of the $x$-dependence (e.g.\ with machine learning).
Obviously, an important direction is also to analyze and quantify all systematic effects at various stages of the multi-step procedure of extracting physical GPDs.
The data can also be used to extract moments of GPDs based on an operator product expansion and as input to the pseudo-distribution approach of Radyushkin \cite{Radyushkin:2019owq}.
Moreover, the approach can be generalized to helicity and transversity GPDs (see Refs.~\cite{Alexandrou:2020zbe,Alexandrou:2021bbo} for our results in the symmetric frame), as well as to twist-3 GPDs \cite{Bhattacharya:2021oyr}.

Overall, our approach offers new possibilities for mapping out the whole kinematic dependence of GPDs and obtaining precise tomographic pictures of hadrons. As such, it can play an important complementary role to experimental, phenomenological and other theoretical analyses and contribute to the quest of achieving profound understanding of the fundamental properties of the nucleon and other hadrons.

\vspace*{-2mm}
\begin{scriptsize}
\section*{Acknowledgments}
KC\ is supported by the grants SONATA BIS No.\ 2016/22/E/ST2/00013 and OPUS No.\ 2021/43/B/ST2/00497 (National Science Centre, Poland).
SB and SM are supported by the U.S. Department of Energy (DoE), Office of Science (OoS), Office of Nuclear Physics (OoNP) through Contract No.~DE-SC0012704, No.~DE-AC02-06CH11357 and within the framework of Scientific Discovery through Advance Computing (SciDAC) award Fundamental Nuclear Physics at the Exascale and Beyond.
MC, JD and AS acknowledge financial support from DoE, OoNP, Early Career Award under Grant No.\ DE-SC0020405.
JD also received support by the DoE, OoS, OoNP, within the framework of the TMD Topical Collaboration. 
AM was supported by the National Science Foundation under grant No.\ PHY-2110472, and also by DoE, OoS, OoNP, within the framework of the TMD Topical Collaboration. 
FS was funded by the NSFC and the Deutsche Forschungsgemeinschaft (DFG, German Research Foundation) through the funds provided to the Sino-German Collaborative Research Center TRR110 “Symmetries and the Emergence of Structure in QCD” (NSFC Grant No.\ 12070131001, DFG Project-ID 196253076 - TRR 110). 
YZ was partially supported by an LDRD initiative at Argonne National Laboratory under Project~No.~2020-0020.
Computations for this work were carried out on facilities of: the USQCD Collaboration funded by OoS of DoE, the Oak Ridge Leadership Computing Facility, which is a
DOE OoS User Facility supported under Contract DE-AC05-00OR22725, PLGrid Infrastructure (Prometheus at AGH Cyfronet in Cracow), Poznan Supercomputing and Networking Center (Eagle), Interdisciplinary Centre for Mathematical and Computational Modelling of the Warsaw University (Okeanos), Academic Computer Centre in Gda\'nsk (Tryton).
The gauge configurations were generated by the Extended Twisted Mass Collaboration on the KNL (A2) Partition of Marconi at CINECA, through the Prace project Pra13\_3304 ``SIMPHYS".
Inversions were performed using the DD-$\alpha$AMG solver~\cite{Frommer:2013fsa} with twisted mass support~\cite{Alexandrou:2016izb}. \vspace*{-5mm}
\end{scriptsize}

\bibliographystyle{h-physrev}
\bibliography{references}

\end{document}